\documentclass[aps,prl,10pt,twocolumn,superscriptaddress,longbibliography]{revtex4}
\usepackage{graphicx}
\usepackage{amssymb, amsmath,amsfonts}
\usepackage{color}
\usepackage{ulem}
\usepackage{bbm}
\usepackage{subfigure}
\usepackage{dcolumn}
\usepackage{mathtools}
\usepackage{pifont}
\usepackage{MnSymbol}
\usepackage{pdfpages}

\def\I{\uppercase\expandafter{\romannumeral 1}}
\def\II{\uppercase\expandafter{\romannumeral 2}}
\def\III{{\uppercase\expandafter{\romannumeral 3}}}
\def\IV{{\uppercase\expandafter{\romannumeral 4}}}
\def\V{{\uppercase\expandafter{\romannumeral 5}}}
\def\VI{{\uppercase\expandafter{\romannumeral 6}}}
\def\VII{{\uppercase\expandafter{\romannumeral 7}}}
\def\i{\lowercase\expandafter{\romannumeral 1}}
\def\ii{\lowercase\expandafter{\romannumeral 2}}
\def\iii{{\lowercase\expandafter{\romannumeral 3}}}
\def\iv{{\lowercase\expandafter{\romannumeral 4}}}
\def\v{{\lowercase\expandafter{\romannumeral 5}}}
\def\vi{{\lowercase\expandafter{\romannumeral 6}}}
\def\vii{{\lowercase\expandafter{\romannumeral 7}}}
\def\nn{\nonumber\\}

\begin{document}
%\linenumbers

\title{Theory for charge density wave and orbital-flux state in antiferromagnetic kagome metal FeGe}

\author{Hai-Yang Ma}
\affiliation{School of Physical Science and Technology, ShanghaiTech University, Shanghai 201210, China}
\affiliation{ShanghaiTech laboratory for topological physics, ShanghaiTech University, Shanghai 201210, China}

\author{Jia-Xin Yin}
\affiliation{Laboratory for Topological Quantum Matter and Advanced Spectroscopy (B7), Department of Physics, Princeton University, Princeton,
New Jersey 08544, USA}

\author{M. Zahid Hasan}
\affiliation{Laboratory for Topological Quantum Matter and Advanced Spectroscopy (B7), Department of Physics, Princeton University, Princeton,
New Jersey 08544, USA}
\affiliation{Princeton Institute for Science and Technology of Materials, Princeton University, Princeton, New Jersey 08544, USA}
\affiliation{Materials Sciences Division, Lawrence Berkeley National Laboratory, Berkeley, California 94720, USA}

\author{Jianpeng Liu}
\email[]{liujp@shanghaitech.edu.cn}
\affiliation{School of Physical Science and Technology, ShanghaiTech University, Shanghai 201210, China}
\affiliation{ShanghaiTech laboratory for topological physics, ShanghaiTech University, Shanghai 201210, China}

\begin{abstract}
In this work, we theoretically study the charge order and orbital magnetic properties of a new type of antiferromagnetic kagome metal FeGe.  
Based on first principles density functional theory (DFT) calculations, we have studied the electronic structures, Fermi-surface quantum fluctuations, as well as phonon properties of the antiferromagnetic kagome metal FeGe. We find that charge density wave emerges in such a system due to a subtle cooperation between electron-electron ($e$-$e$) interactions and electron-phonon couplings, which gives rise to an unusual scenario of interaction-triggered phonon instabilities, and eventually yields a charge density wave (CDW) state. 
We further show that, in the CDW phase,  the ground-state current density distribution exhibits an intriguing star-of-David pattern, leading to flux density modulation. The orbital fluxes (or current loops) in this system emerges as a result of the subtle interplay between magnetism, lattice geometries, charge order, and spin-orbit coupling (SOC), which can be described by a simple, yet universal, tight-binding theory including a Kane-Mele type SOC term and a magnetic exchange interaction. We further study the origin of the peculiar step-edge states in FeGe, which shed light on the topological properties and correlation effects in this new type of kagome antiferromagnetic material.   
\end{abstract}


\maketitle

%\section{Introduction}
The frustrated nature of a kagome lattice  can give rise to intriguing electronic structure such as flat bands, Dirac cones, and van Hove singularities (vHS) \cite{balents-prb08,Zhou2017, Neupert2022,yin-review-nature22}.  The flat bands may serve as a good platform for strongly correlated physics such as quantum magnetism and unconventional superconductivity \cite{wen-prl11,thomale-kagome-prl13,wang-kagome-prb13,thomale-kagome-prb12,li-kagome-prb12,yin2018giant,ye2018massive,liu2018giant,yin2020quantum,ko2009doped,balents2010spin,kiesel2013unconventional}. The Dirac points can be gapped out either by  spontaneous time-reversal symmetry breaking  or spin-orbit coupling (SOC), which thus exhibits nontrivial topological properties \cite{Yin2018, Ye2018, Xu2015}.  When the Fermi level is tuned to the van Hove singularity of a kagome metal, various correlated states may emerge through the Fermi-surface nesting scenario as driven by $e$-$e$ Coulomb interactions \cite{wu2021nature,feng2021chiral,gu2021gapless,balents-instability-prb21,Ortiz2020,Jiang2021,Zhao2021,Ortiz2021,Chen2021,ChenH2021,Yu2021,Xu2021,Yin2021,Li2021,Zhang2021,Ortiz2021PRX,Wu2021,Kang2021,yu2021evidence,Nakayama2021,Li2022,Liu2021,Cho2021,Zhou2021,Xie2022,Wulferding2021,Wang2021,Song2021,Nie2022,mielke-muon-nature22,Lou2022,Li2022np,Wu2022,shan-muon-prr22}.

Recently, scanning tunnelling microscopy (STM) and muon spin relaxation ($\mu$sR) measurements reveal time-reversal breaking charge orders in AV$_3$Sb$_5$ (A=K, Rb, Cs) \cite{jiang2021unconventional,mielke-muon-nature22, yu2021evidence,shan-muon-prr22}, which emerge at temperatures near or below the $2\times 2$ CDW transitions for this class of materials.  
Interestingly, these time-reversal breaking CDW phases are proposed to exhibit spontaneous real-space current loops, and are associated with orbital flux patterns \cite{varma-prb97, liu-prl16,liu-nrp21,bourges-review21,feng2021chiral, denner-prl21, Ma2021}. Moreover, recent experimental studies demonstrate that an antiferromagnetic metal FeGe with kagome lattice structure also hosts a 2$\times$2  charge order \cite{yin-arxiv22,teng2022discovery}, which is closely correlated with the antiferromagnetism \cite{fege-arpes-arxiv22,lee-fege-arxiv22}.  Despite different theoretical proposals \cite{chang-fege-arxiv22,wan-fege-dmi-arxiv22,wan-fege-cdw-arxiv22}, the origin of CDW state and how the charge order interplays with antiferromagnetism in FeGe are still open questions. 
In this work we theoretically study the properties of charge density wave and the orbital-flux states in the antiferromagnetic kagome metal FeGe. 
Based on first principles DFT calculations, we have systematically studied the electronic structures, Fermi-surface quantum fluctuations, and phonon properties of FeGe. Specifically, through generalized susceptibility calculations within random phase approximation (RPA), we find  that $e$-$e$ interactions along cannot drive any CDW-type instability in FeGe. However, $e$-$e$ interactions would renormalize the effective mass around the Fermi surface which thus  enhances the density of states contributed by Ge $p$ orbitals. The latter is coupled with a branch of low-frequency optical phonon modes around $M$ point, which are consisted of Ge atomic displacements.
This branch of Ge phonon modes is getting softer with the increase of $U$ as more and more Ge $4p$ orbitals contribute to the Fermi surface,  and eventually gets frozen with the onset of CDW. Therefore, we propose that the CDW state in FeGe is triggered  by an interaction-assisted phonon instability.

We further show that, in the CDW phase,  the inter-site currents in the ground state form a mutually intercalated star-of-David and double-triangular current pattern. Such orbital fluxes (or current loops) in this system emerges as a result of the subtle interplay between magnetism, lattice geometries, charge order, and spin-orbit coupling (SOC), which can be described by a simple, yet universal, tight-binding model including a Kane-Mele type SOC term and a magnetic exchange interaction. 
Moreover, we have also studied FeGe in the monolayer form  in the $2\times 2$ CDW phase, which turns out to be a topological metal with both Weyl points and flat Chern band near the Fermi level, exhibiting multiple low-energy edge states. This naturally explains the origin of the recently observed spin polarized step-edge states in FeGe.

\begin{figure}[!htbp]
\includegraphics[width=0.48\textwidth]{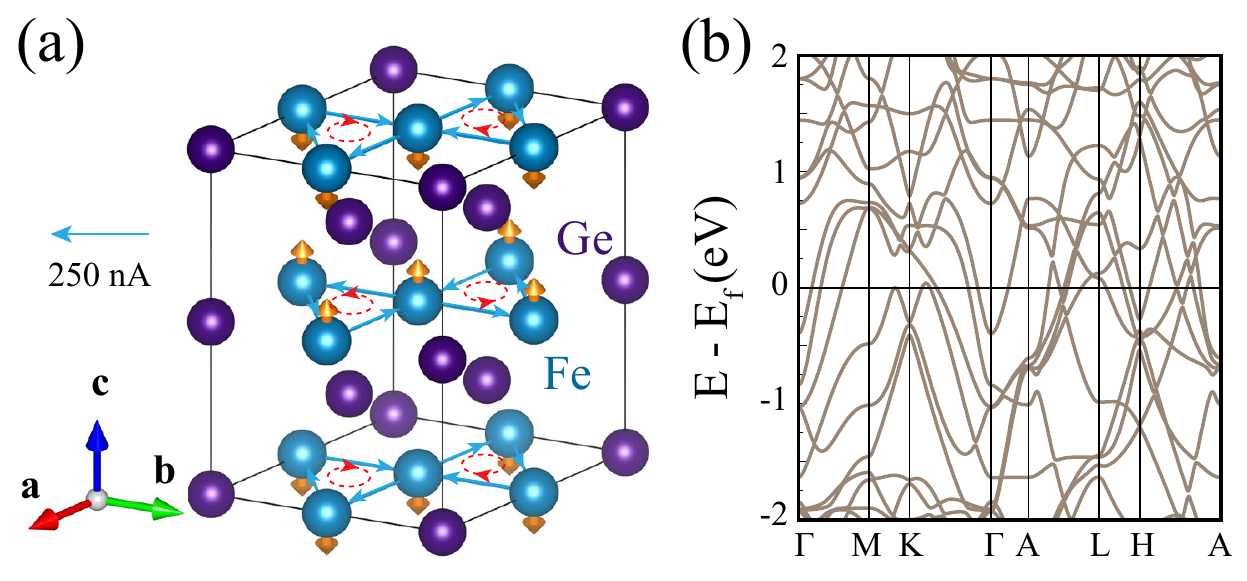}
\caption{~\label{fig1} Lattice structure, current patterns and band structures of pristice FeGe. (a) Lattice structure and current patterns. The inter-site current flow are indicated by solid arrows, while the current loops among the Fe triangles are illustrated by the arrowed dash circles. (b) Band structures.}
\end{figure}

\begin{figure*}[!htbp]
\includegraphics[width=1.0\textwidth]{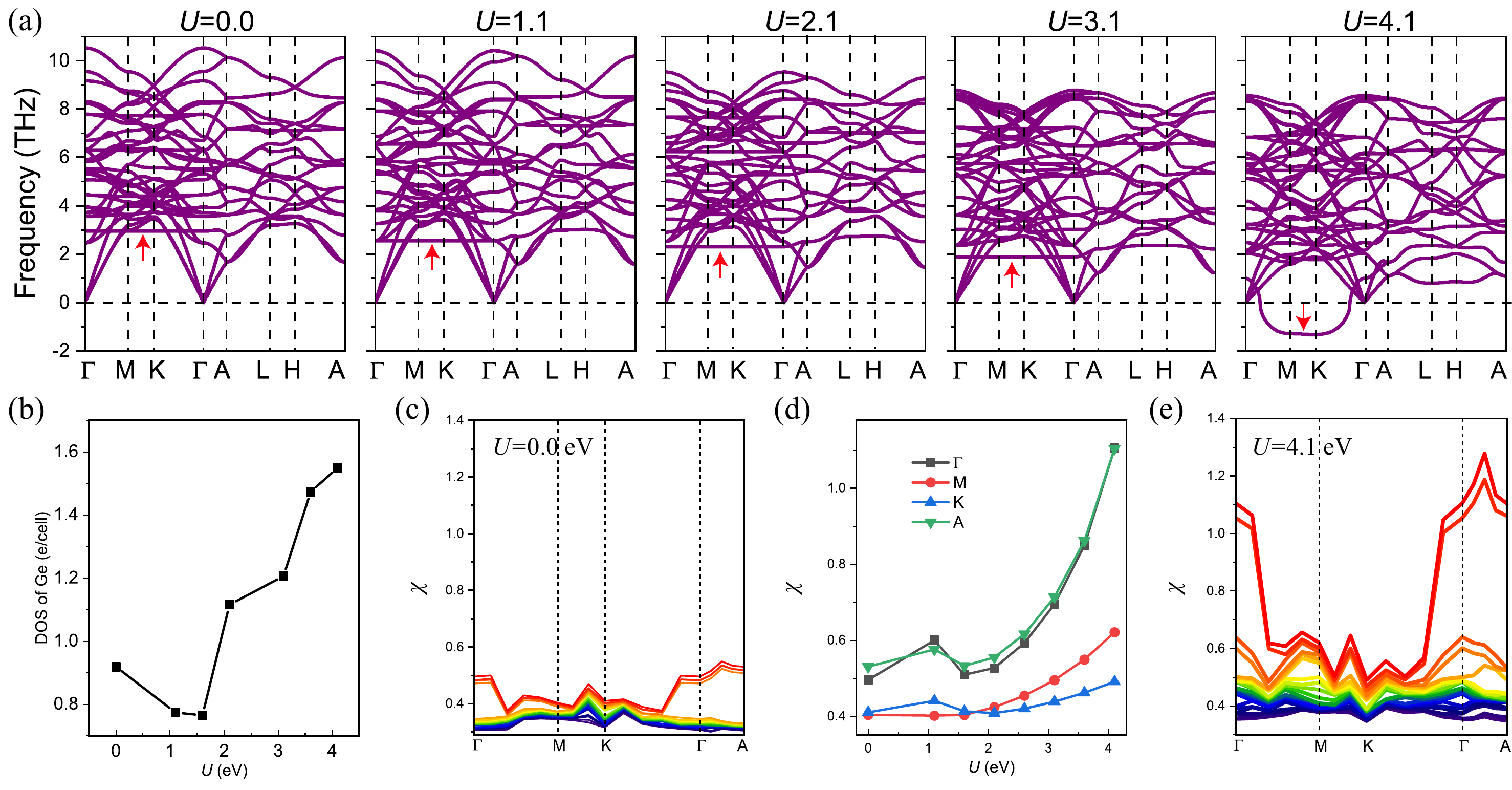}
\caption{~\label{fig2} (a) Phonon spectrums. (b) Dependence of the DOS of the Ge on the on-site $U$, taken at the Fermi energy. (c) Leading eigenvalues of the susceptibility tensors along a high-symmetry \textbf{q} path, taken at $U=0.0$~eV. (c) Leading eigenvalues of the RPA susceptibility tensors of pristine FeGe at different Hubbard $U$ values. (d) Same at (b), but taken at $U=4.1$~eV. The flat bands in (a) which goes down with increasing $U$ are indicated by red arrows. Here $J=0$~eV for $U=0$~eV and $J=0.82$~eV otherwise.}
\end{figure*}

%\subsection{Pristine structure of FeGe}
FeGe consists of an alternating stacking of the Fe$_3$Ge kagome layer and the Ge honeycomb layer as shown in Fig.~\ref{fig1}(a). 
It has an $A$-type collinear antiferromagnetic order along the $c$ axis which doubles the primitive cell with a N\'eel temperature $\sim 400\,$K \cite{Bernhardt1984}. 
We perform DFT calculations with additional Hubbard-like on-site interactions applied to the Fe $3d$ orbitals \cite{Liechtenstein1995}, where the Hubbard $U$ is treated as a parameter to be varied. With $U=4.1\,$eV (determined through the linear-response method \cite{Cococcioni2005}) and Hund's coupling $J=U/5=0.82\,$eV, the calculated local magnetic moment for each Fe atom is 2.61\,$\mu_{\rm{B}}$. The band structures in the antiferromagnetic phase of the pristine structure are shown in Fig.~\ref{fig1}(b), which involve multiple bands around the Fermi level. Moreover, these low-energy bands are topologically nontrivial, which host multiple  Dirac nodal loops as marked by the green circles in Fig.~S5 in Supplementary Information (SI). 
The nodal loops are resided within the $k_z\!=\!\pi$ plane, which are protected by horizontal mirror symmetry \citep{supp_info}.
More details about the band topology are given in SI \cite{supp_info}.

\begin{figure*}[tbh]
\includegraphics[width=0.75\textwidth]{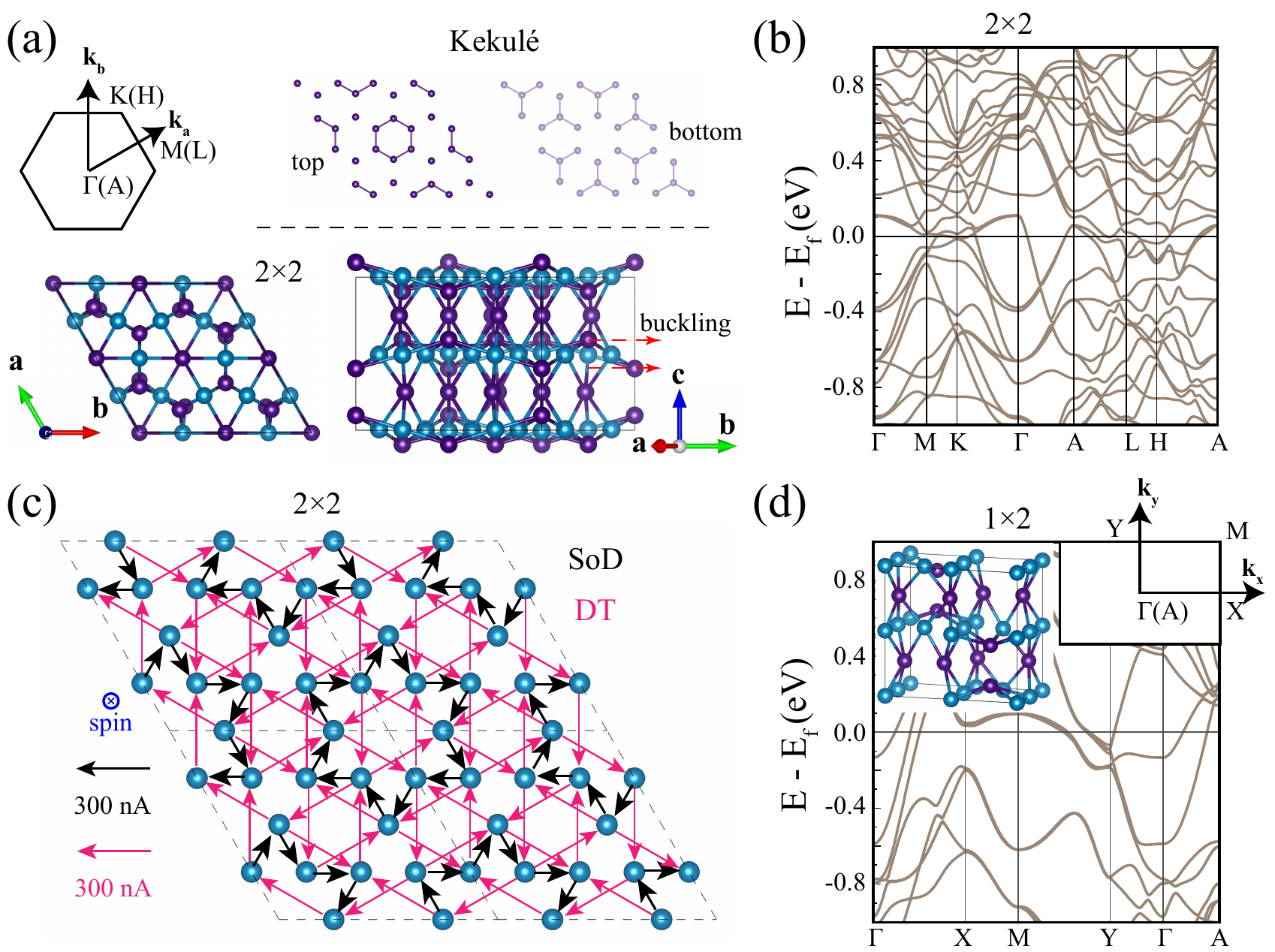}
\caption{~\label{fig3} Lattice, electronic structures and current-loop states of the CDW phase of FeGe. (a) First BZ, relaxed lattice structure, illustration of the kekul\'e structure, and buckling of the 2$\times$2 CDW structure of FeGe. The kekul\'e structure is formed by the distortion of the Ge atoms in the Ge planes. The bottom-right panel is the side view to show the buckling of the Ge atoms in the kagome plane. (b) Band structures of the 2$\times$2 supercell structure. (c) Current patterns of the 2$\times$2 supercell structure. SoD is the abbreviation for star of David and DT is the abbreviation for double triangular, see the main text for details. (d) Relaxed lattice structure and band structures of the 1$\times$2 supercell structure.}
\end{figure*}

We continue to discuss the origin of the CDW phase in FeGe. Recent experiments report an in-plane 2$\times$2  charge order at temperatures below $\sim 100\,$K  in FeGe, which seems to be interwined with its antiferromagnetic order \cite{yin-arxiv22}. 
To shed light on the experiments, we first calculate the phonon spectra of pristine FeGe at different Hubbard $U$ values, the results are shown in Fig.~\ref{fig2}(a). As we can see, at small $U$, the phonon spectrum of the pristine structure are well behaved without any imaginary frequency, indicating that the pristine structure is stable. With the increase of $U$ value, a flat phonon band within the $k_z=0$ plane, as marked by red arrow in Fig.~\ref{fig2}(a), gradually moves downward in frequency and  becomes imaginary when $U>3.1\,$eV, indicating that the pristine structure eventually gets unstable at larger $U$ values, possibly with the onset of CDW. A closer inspection reveals that the soft branch of phonon modes are contributed by collective displacements of the Ge atoms. This unusual dependence of the phonon spectrum on the Hubbard $U$ implies that the CDW transition in FeGe may be triggered by $e$-$e$ interactions. %To better understand the mechanism behind such a scenario of interaction-assisted CDW transition, we need to look more carefully at the electronic band structures at different $U$ values. 
It turns out that with the increase of $U$, the flat bands contributed from the Fe 3$d$ orbitals are gradually pushed away from the Fermi level (see Supplementary Information); in the meanwhile, contributions from the Ge $4p$ orbitals at the Fermi level become more and more significant (Fig.~\ref{fig2}(b)) with the increase of $U$. As the soft phonon band (marked in Fig.~\ref{fig2}(a)) is contributed by Ge displacements, it is coupled to the Ge 4$p$ orbitals much  more strongly than to the Fe $3d$ orbitals. Thus, an enhancement of the Ge $4p$ spectral weight at the Fermi surface would yield stronger electron-phonon coupling matrix elements for the soft phonon band with the Ge displacements. This would give rise to  a strong renormalization effects to the phonon frequency \cite{epi-rmp17}, and eventually makes the soft phonon band (from Ge displacements) unstable, driving the CDW transition. Such an argument is also consistent with the relaxed structure of the $2\times 2$ CDW phase as shown in Fig.~\ref{fig3}(a), in which only the Ge atoms show substantial distortions. We will discuss this in greater details later. 

We note that the Hubbard $U$ value  can be calculated self consistently through the response process of local charge density to a perturbative potential \cite{Cococcioni2005}, which gives $U=4.1\,$eV for Fe $3d$ orbitals. This justifies the above argument that CDW of FeGe may be induced through an interaction-assisted phonon instability scenario. Certainly  DFT+$U$  is  a mean-field approach which cannot fully capture the characteristics of single-particle excitation spectrum for such correlated metals \. However, recently both angle resolved photoemission spectroscopy (ARPES) measurements \cite{fege-arpes-arxiv22} and DFT + dynamical mean field theory (DMFT) calculations \cite{lee-fege-arxiv22} suggest an overall effective mass renormalization by a factor of $\sim$1.6 compared to DFT band structures, with well defined quasi-particle features around Fermi level. As the Fermi surface has contributions from both Fe $3d$ and Ge $4p$ orbitals,  both the overall density of states (DOS) and Ge $4p$ DOS around the Fermi level would be enhanced due to $e$-$e$ interaction effects. This further confirms the above scenario that interaction-enhanced DOS would lead to stronger electron-phonon coupling effects thus induces phonon instability.

We also consider the possibility that the CDW state in antiferromagnetic FeGe may be driven by Fermi-surface instabilities due to strong $e$-$e$ interaction effects.  We calculate the general susceptibility tensor defined in the charge-sublattice-orbital-spin space \cite{liu-prb17,motome-prb15,xu-prl21,ma-prb22} based on a realistic Wannier tight binding model including all the Fe $3d$ orbitals and Ge $4p$ orbitals generated from DFT calculations. The eigenmodes and eigenvalues of the static susceptibility tensor reflect the properties of the intrinsic quantum fluctuations at the Fermi surface. The $e$-$e$ interaction renormalization effects on the generalized susceptibility tensor are treated by random phase approximation (RPA). The leading eigenmodes of the RPA generalized susceptibility tensor indicate the possible spontaneous symmetry breaking states driven by Fermi surface quantum fluctuations and $e$-$e$ interactions. The details are given in Supplementary Information.
%to find the possible Fermi surface instabilities. 
In Fig.~\ref{fig2}(d), we show the  eigenvalues of  the bare susceptibility tensor calculated from the $U=0$ Fermi surface in the antiferromagnetic ground state of FeGe with pristine lattice structure. We see that  the eigenvalues of all the leading Fermi-surface fluctuation modes are small, with amplitudes $\lessapprox 0.5$. Including $e$-$e$ interactions in the RPA framework does not change the result qualitatively. As shown in Fig.~\ref{fig2}(e), the largest eigenvalues of the RPA susceptibility tensor at different high-symmetry points ($\Gamma$, $M$, $K$, $A$) only show moderate enhancement with the increase of $U$, which are far from driving a CDW transition (marked by diverging susceptibility eigenvalue). We have further inspected the effects of inter-site Coulomb interactions (between Fe $3d$ and Ge $4p$ electrons),  and still cannot find any instability mode that can lead to CDW state (see supplementary information). 
Therefore, we conclude that in FeGe $e$-$e$ interactions alone cannot lead to a CDW-type Fermi surface instability; rather the CDW state results from a subtle interplay between on-site $e$-$e$ interactions and electron-phonon couplings, which realizes an intriguing scenario of interaction-assisted phonon instability, giving rise to the CDW state.

As the phonon spectrum at $U=4.1\,$eV has a whole optical-phonon branch with imaginary frequencies within the $k_z\!=\!0$ plane (Fig.~\ref{fig2}), which implies that there could be various possible lattice distortions. Here we only consider the unstable phonon modes at the three $M$ points in the pristine Brillouin zone, which are the modes that first become unstable as $U$ increases. We first make a linear combination of the three unstable phonon modes at the three $M$ points, which yield a $2\times 2$ supercell structure. Then we take such a lattice distortion as an initial ansatz for the structural relaxation calculation.  
In Fig.~\ref{fig3}(a) we present the fully relaxed lattice structure of a $2\times 2$ supercell, which preserves all the symmetries of the pristine structure. This is a stable structure as no more imaginary frequency can be found in its phonon spectrum as shown in Fig.~S6 in SI \cite{supp_info}. Such a $2\times 2$ CDW phase involves two types of lattice distortions. The first one is  a type of in-plane shortening or elongation of the nearest neighbour Ge-Ge bond within the honeycomb Ge layer, which forms a ``kekul\'e type" distortion pattern of the Ge honeycomb lattice as shown in the top panel of Fig.~\ref{fig3}(a), where the two Ge atoms connected by a solid line denotes the shortened Ge-Ge bond, consistent with previous report \cite{chang-fege-arxiv22}. The kekul\'e distortions of the two Ge honeycomb layers within an antiferromagnetic primitive cell stacks along the $z$ direction in an anti-phase way as illustrated by the opposite kekul\'e distortions for the bottom and top Ge layers [top panel of Fig.~\ref{fig3}(a)]. The second type of distortion is the out-of-plane buckling of the Ge atoms within the Fe kagome plane, as shown in the bottom panel of Fig.~\ref{fig3}(a). We have also considered the situation that only one of the three unstable phonon modes at the $M$ points is stablized and leads to a CDW phase with $2\times 1$ supercell, whose relaxed lattice structure  is shown in the inset of Fig.~\ref{fig3}(d). Although the $1\times 2$ supercell is slightly lower in energy than the $2\times 2$ one, the former is inconsistent with recent experimental observations \cite{yin-arxiv22,dai-fege-arxiv22}. Thus, the CDW phase observed in experiments may be the $2\times 2$ supercell resulted from the spontaneous condensation of three unstable phonon modes at the three $M$ points, which are characterized by in-plane kekul\'e distortions and out-of-plane bucklings of the Ge atoms.

We then turn to the discussion of the electronic structures and orbital magnetic properties of antiferromagnetic FeGe within the CDW phase. In Fig.~\ref{fig3}(b) we present the band structures of the antiferromagnetic FeGe in the $2\times 2$ lattice structure, where every band is twofold degenerate due to the combined time-reversal ($\mathcal{T}$) and inversion symmetry of the antiferromagnetic phase.%, and there are multiple band crossing the Fermi level, which yields rather complicated Fermi surfaces. 
As mentioned above, a realistic Wannier tight-binding model including all the Fe $3d$ orbitals and Ge $4p$ orbitals has been constructed for the $2\times 2$ antiferromagnetic CDW phase, based on which the inter-site currents have been calculated. As denoted by the arrows in Fig.~\ref{fig3}(c), the current pattern of the Fe kagome layer with spin up magnetization in the $2\times 2$ CDW phase is qualitatively different from that of the pristine phase (Fig.~\ref{fig1}(a)). In the pristine phase of the antiferromagnetic state, the currents flow around a loop connecting the nearest-neighbor Fe triangle within the kagome plane, with the current amplitude $\sim 250\,$nA. In the $2\times 2$ CDW phase,
 as a result of the lattice distortion, the current pattern  within each Fe kagome layer consists of  two types of current loops: the first one only flows between the nearest neighbor Fe sites, forming a star of David (SoD) current loop flowing in the clockwise direction (black arrows in Fig.~\ref{fig3}(c)) with the current amplitude $\sim$360\,nA; the second type only flows between the second nearest neighbor Fe sites, forming double triangular shaped (DT) current loops flowing in the counter-clockwise direction (red arrows in Fig.~\ref{fig3}(c)) with the current amplitude $\sim$280\,nA. These two types of current loops intersect with each other, forming an intriguing ``current-loop density wave state" that is driven by the subtle interplay among magnetism, spin-orbit coupling (SOC), and the charge order. The current pattern in the Fe kagome layer with spin down magnetization is exactly opposite to that of the spin-up layer. 
This forms an new intralayer ferromagnetic and interlayer antiferromagnetic orbital magnetic order that can be potentially measured by  neutron diffraction measurements.

The current-loop state in FeGe can be captured by a simple tight-binding model. For a monolayer ferromagnetic kagome lattice, the tight-binding Hamiltonian can be written as :
\begin{align}
\hat{H}_{\rm{mono}}=&-t\sum_{\langle i j\rangle}\hat{c}^{\dagger}_{i}\hat{c}_{j}+i\lambda\sum_{\langle i\alpha,j\beta\rangle}(\mathbf{E}_{ij}\times\mathbf{r}_{ij})_{z}\cdot\sigma_{z}^{\alpha\beta}\hat{c}^{\dagger}_{i\alpha}\hat{c}_{j\beta}\nn
&+J\sum_{i\alpha\beta}\mathbf{S}_i\cdot\mathbf{\sigma}^{\alpha\beta}\hat{c}^{\dagger}_{i\alpha}\hat{c}_{i\beta}
\label{eq:Hmono}
\end{align}
where $\langle\rangle$ refers to the nearest neighbor hopping, $\hat{c}^{\dagger}_{i}\ (\hat{c}_{j})$ is the electron creation (annihilation) operator at site $i\ (j)$, $\alpha\beta$ is the spin index. %Similar Hamiltonian was considered in ref. \cite{Zhang2011} for the discussions of the quantum anomalous Hall effect in monolayer ferromagnetic kagome lattice. 
The first term is the usual first-neighbor hopping within the kagome triangles. The second term is the Kane-Mele type SOC term \cite{Kane2005} which is generated by the internal in-plane electric field $\mathbf{E}_{ij}$ normal to the bond vector $\mathbf{r}_{ij}$ connecting the two sites $i$ and $j$ [see Supplementary Information]. 
The third term is the effective magnetic exchange interaction. If one periodically stacks the monolayer model along the $z$ direction in an antiferromagnetic configuration with some proper interlayer coupling included, one would obtain a minimal model describing the electronic properties of antiferromagnetic FeGe in the pristine structure. 
Surprisingly, despite the complexity of the Fermi surface topology in the pristine structure of FeGe (Fig.~\ref{fig1}(b) and Fig.~\ref{fig2}(b)), the current pattern from first principles calculations (Fig.~\ref{fig1}(a)) is perfectly consistent with that obtained from the simple tight-binding theory described above.
In general, more terms can be included into the model for a more  realistic description of the system. For example, next-nearest-neighbor hoppings and SOC can be added, which can lead to the DT current pattern; and an effective charge order can also be added to the tight-binding model which give rises to the SoD current pattern. More details about the extensions of the tight-binding model can be found in SI \cite{supp_info}.

\begin{figure}[!htb]
\includegraphics[width=0.48\textwidth]{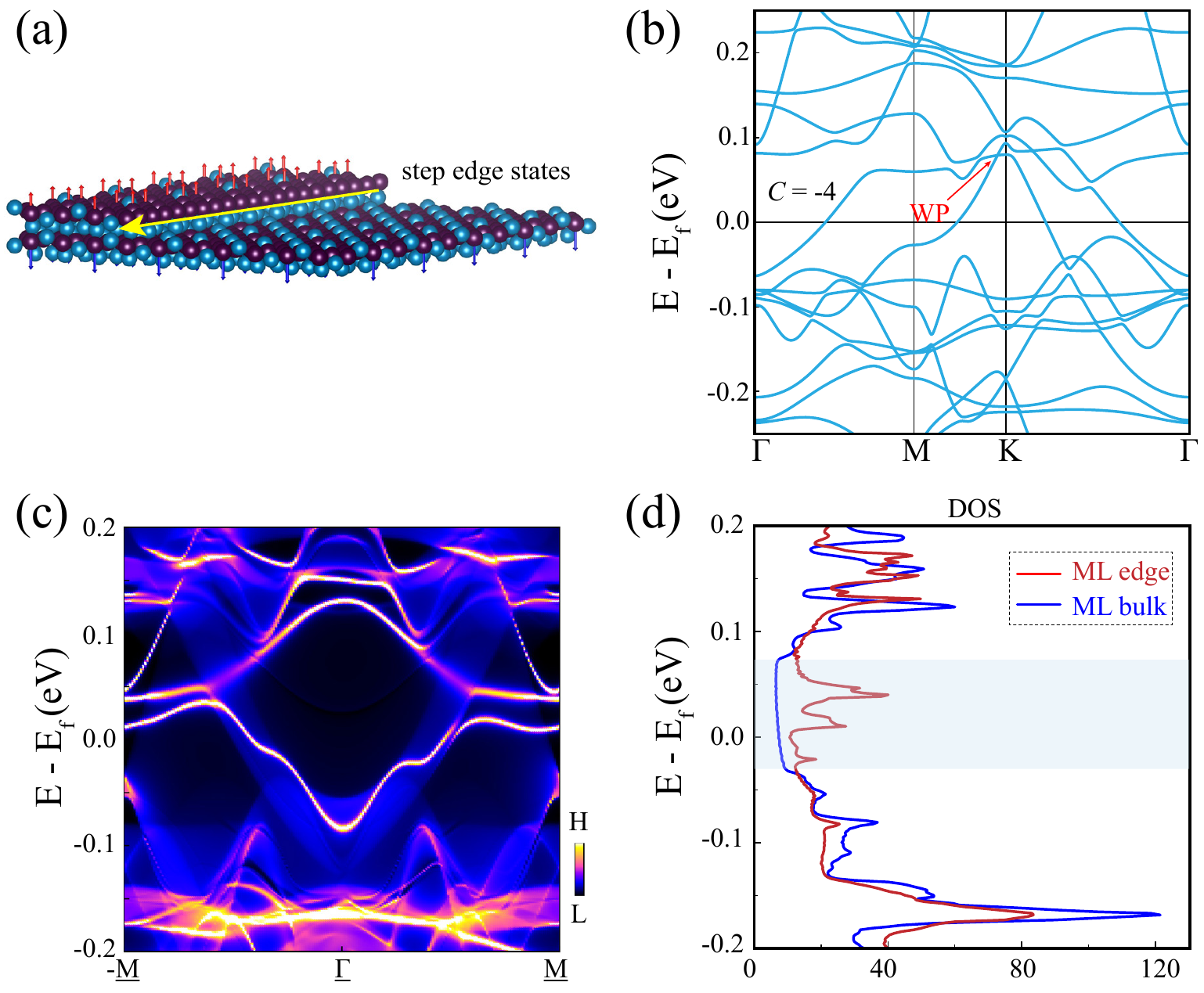}
\caption{~\label{fig4} Topological properties of the single kagome layer of the 2$\times$2 supercell structure of FeGe. (a) Schematics of the step edge states of the 2$\times$2 supercell structure of FeGe. (b) Band structures of monolayer FeGe. The Weyl point protected by $M_x\mathcal{T}$ symmetry is marked by ``WP" along the $K\rm{-}M$ line. (c) Edge states. (d) Density of states (DOS) for the bulk and the edge of the single layer. ``ML bulk" in (d) refer to the DOS of the bulk states of single kagome layer. ML edge refer to DOS of the edge states. The shading area in (d) refer to the gap region of the ML bulk DOS.}
\end{figure}

We note that in experiments, edge states [Fig.~\ref{fig4}(a)] are observed when the bias voltage is within the gap range of the CDW structures \cite{yin-arxiv22}, indicating the topological  nature of the antiferromagnetic CDW phase. To shed light on these observations, we calculate the band structures and edge states of the single kagome layer for the 2$\times$2 structure. The band structures of the monolayer are shown in Fig.~\ref{fig4}(b). We find that there exists both Weyl points (protected by $M_x\mathcal{T}$ symmetry) and topologically nontrival Chern band near the Fermi level, which may lead to nontrivial edge states. As we can see from Fig.~\ref{fig4}(c) that there exists prominent edge states for the single kagome layer. 
We further calculate both the edge and bulk density of states (DOS) for the case of single kagome layer in Fig.~\ref{fig4}(d). A gap feature can be clearly seen in the bulk DOS, while large edge states appear within the  bulk (partial) gap, consistent with the experiments \cite{yin-arxiv22}. More details for the single kagome-layer calculations can be found in Fig.~S8 of SI \cite{supp_info}. Moreover, we note that the Chern band of single-layer FeGe has a small bandwidth ($\sim 0.15\,$eV) with van Hove singularities near the $M$ point near Fermi level [see Fig.~\ref{fig4}(b)], which is expected to  be unstable against electron-electron interactions, and may lead to novel correlated and topological states of matter such as topological density-wave states \cite{young-np22,pablo-np21} and chiral superconductivity \cite{qi-chiral-prb10}.

%\Blue{[JPL: STOP HERE]}

%\section{Conclusion}
%To conclude, in this work we have theoretically studied the mechanism of the charge density wave state in time-reversal breaking antiferromagnetic kagome metal FeGe. states in both the CDW and pristine structures of  several kagome magnetic systems, including FeGe, Co$_3$Sn$_2$S$_2$, Fe$_3$Sn etc. We show that the orbital fluxes in these systems emerge as a result of the subtle interplay between magnetism, lattice geometries, and spin-orbit coupling (SOC), which can be described by a  universal tight-binding theory including both Kane-Mele type SOC term and a magnetic exchange coupling term. 
%We find  that the inter-site currents in the CDW phase of FeGe form a mutually intercalated star-of-David and double-triangular current pattern, which can be potentially measured by neutron diffraction measurements. The monolayer form of FeGe in the CDW phase turns out to be a topological metal hosting both symmetry-protected Weyl points and flat Chern bands, which exhibits multiple low-energy edge states,  which  naturally explains the origin of the recently observed spin polarized step-edge states in FeGe. The flat Chern bands in monolayer FeGe may lead to novel topological and correlated quantum states under proper fillings, which may stimulate further experimental and theoretical studies. Our work sheds light on the intriguing magnetic and charge properties in kagome magnets, and will provide useful guidelines for future works.

To conclude, based on first principles  calculations, we have systematically studied the electronic structures, Fermi-surface quantum fluctuations, as well as phonon properties of the antiferromagnetic kagome metal FeGe. We find that charge density wave emerges due to a subtle cooperation between $e$-$e$ interactions and electron-phonon couplings, which leads to an unusual scenario of interaction-triggered phonon instabilities, and eventually yields a charge density wave (CDW) state. 
We further show that, in the CDW phase of such antiferromagnetic kagome metal,  the ground-state current density distribution exhibits an intriguing star-of-David pattern, leading to flux density modulation, which can be potentially measured by neutron diffraction measurements. Such current loop pattern emerges as a result of the subtle interplay between magnetism, lattice geometry, charge order, and SOC, which can be described by a simple, yet universal, tight-binding theory including a Kane-Mele type SOC term and a magnetic exchange interaction. The monolayer form of FeGe in the CDW phase turns out to be a topological metal hosting both symmetry-protected Weyl points and flat Chern bands, which exhibits multiple low-energy edge states,  which  naturally explains the origin of the recently observed spin polarized step-edge states in FeGe. The flat Chern bands in monolayer FeGe may lead to novel topological and correlated quantum states under proper fillings, which may stimulate further experimental and theoretical studies. Our work sheds light on the intriguing magnetic and charge properties in kagome magnets, and will provide useful guidelines for future works.
%We further study the origin of the peculiar step-edge states in FeGe, which shed light on the topological properties and correlation effects in this new type of kagome antiferromagnetic material.   

\acknowledgements
 We thank Quansheng Wu and Xiangang Wan for valuable discussions. This work is supported by   the National Natural Science Foundation of China (grant No. 12174257),  the National Key R \& D program of China (grant No. 2020YFA0309601), and the start-up grant of ShanghaiTech University. 

\widetext
\clearpage

\begin{center}
\textbf{\large Supplementary Information for ``Theory for charge density wave and orbital-flux state in antiferromagnetic kagome metal FeGe"}
\end{center}

\section{\I. First principles methods}
\subsection{Details for the first-principles calculations}
The first-principles calculations are performed with the Vienna ab initio simulation package (VASP) which adopts the projector-augmented wave method \cite{kresse1996efficient}. The energy cutoff is set at 465 eV for FeGe and 402 eV for Co$_{3}$Sn$_{2}$S$_{2}$, Fe$_{3}$Sn and FeSn. Exchange-correlation functional of the Perdew-Burke-Ernzerhof (PBE) type is used for both the structural relaxations and electronic structures calculations \cite{perdew1996generalized}. The convergence criteria for the total energy is set to 10$^{-6}$ eV. The BZ is sampled by a 10$\times$10$\times$6, 10$\times$10$\times$4, 12$\times$12$\times$16. 10$\times$10$\times$12 $\mathbf{k}$ mesh for the pristine structure of FeGe, Co$_{3}$Sn$_{2}$S$_{2}$, Fe$_{3}$Sn and FeSn , respectively. For the 2$\times$2 and 1$\times$2 supercell structure of FeGe, a 6$\times$6$\times$5 and a 10$\times$6$\times$6 k-meshs are adopted with the $\Gamma$-centered scheme. Rotationally invariant ``DFT+$U$" scheme \cite{Liechtenstein1995} is adopted for some of  the calculations, where the Hubbard on-site $U$ is determined through the linear-response method \cite{Cococcioni2005} for FeGe (with $U=4.1$\,eV).   The Hund's rule coupling is set to $J=U/5$.  We have also calculated the current loop patterns for other magnetic metals including  Fe$_3$Sn ($U=4.3$\,eV) and Co$_{3}$Sn$_{2}$S$_{2}$ (with $U=4.0\,$eV), see Sec.~\ref{sec:material}.

\subsection{Inter-site currents}
Inter-site current patterns are calculated based on the Wannier tight-binding models which are obtained through the VASP2WANNIER90 interface \cite{mostofi2008wannier90}. Here we give the details of the fomalism.
The rate of change of charge density at an atomic site $j$ is expressed as
\begin{equation}
%\begin{aligned}
%\frac{dn_{j}}{dt}=&{\rm Tr}\Big[\frac{d\hat{n}_{j}}{dt}\hat{\rho}\Big]\\
%=&\sum_{\alpha'j'}\langle\phi_{\alpha'j'}|\frac{d\hat{n}_{j}}{dt}\hat{\rho}|\phi_{\alpha'j'}\rangle\\
%=&\frac{i}{\hbar}\sum_{\alpha'j'}\langle\phi_{\alpha'j'}|[\hat{H},\hat{n}_{j}]\,\hat{\rho}\,|\phi_{\alpha'j'}\rangle\\
%=&-\frac{2}{\hbar}{\rm Im}\sum_{\alpha\alpha'j'}\rho_{\alpha j,\alpha'j'}(\mathbf{R})H_{\alpha'j',\alpha j}(-\mathbf{R})\ ,
\frac{dn_{j}}{dt}=-\frac{2}{\hbar}{\rm Im}\sum_{\alpha\alpha'j'}\rho_{\alpha j,\alpha'j'}(\mathbf{R})H_{\alpha'j',\alpha j}(-\mathbf{R})\ ,
%\end{aligned}
\label{eq:inter-site-current}
\end{equation}
where \{$\alpha$, $\alpha'$\} refers to the orbital indices and \{$j$, $j'$\} refers to the site (or sublattice) indices. $\hat{n}_{j}=\sum_{\alpha}|\phi_{\alpha j}\rangle\langle\phi_{\alpha j}|$ is the electron number operator in the Wannier basis, and $\hat{\rho}=\sum_{n,\mathbf{k}}|\psi_{n,\mathbf{k}}\rangle\langle\psi_{n,\mathbf{k}}|\theta(E_{F}-E_{n,\mathbf{k}})$ ($n$ is the band index, $\theta(E_{F}-E_{n,\mathbf{k}})$ is the step function) is the density operator in the Bloch basis. Here $\rho_{\alpha j,\alpha'j'}(\mathbf{R})$ is expressed as
%\begin{equation}
$\rho_{\alpha j,\alpha'j'}(\mathbf{R})=\sum_{\mathbf{k}}e^{-i\mathbf{kR}}\rho_{\alpha j,\alpha'j'}(\mathbf{k})/N$, 
%\end{equation}
and
%\begin{equation}
$H_{\alpha j,\alpha'j'}(\mathbf{-R})=\sum_{\mathbf{k}}e^{i\mathbf{kR}}H_{\alpha j,\alpha'j'}(\mathbf{k})/N$.
%\end{equation}
We define the inter-site current between site $j'$ and $j$ as $I_{jj'}=-(2/\hbar)\rm Im \sum_{\alpha,\alpha'}\,\rho_{\alpha j,\alpha' j'}(\mathbf{R})\,H_{\alpha'j',\alpha j}(-\mathbf{R})$.

\section{\II. Topological properties and electronic structures of pristine FeGe}
\subsection{Topological properties of pristine FeGe}
\begin{figure}[!htbp]
\includegraphics[width=0.7\textwidth]{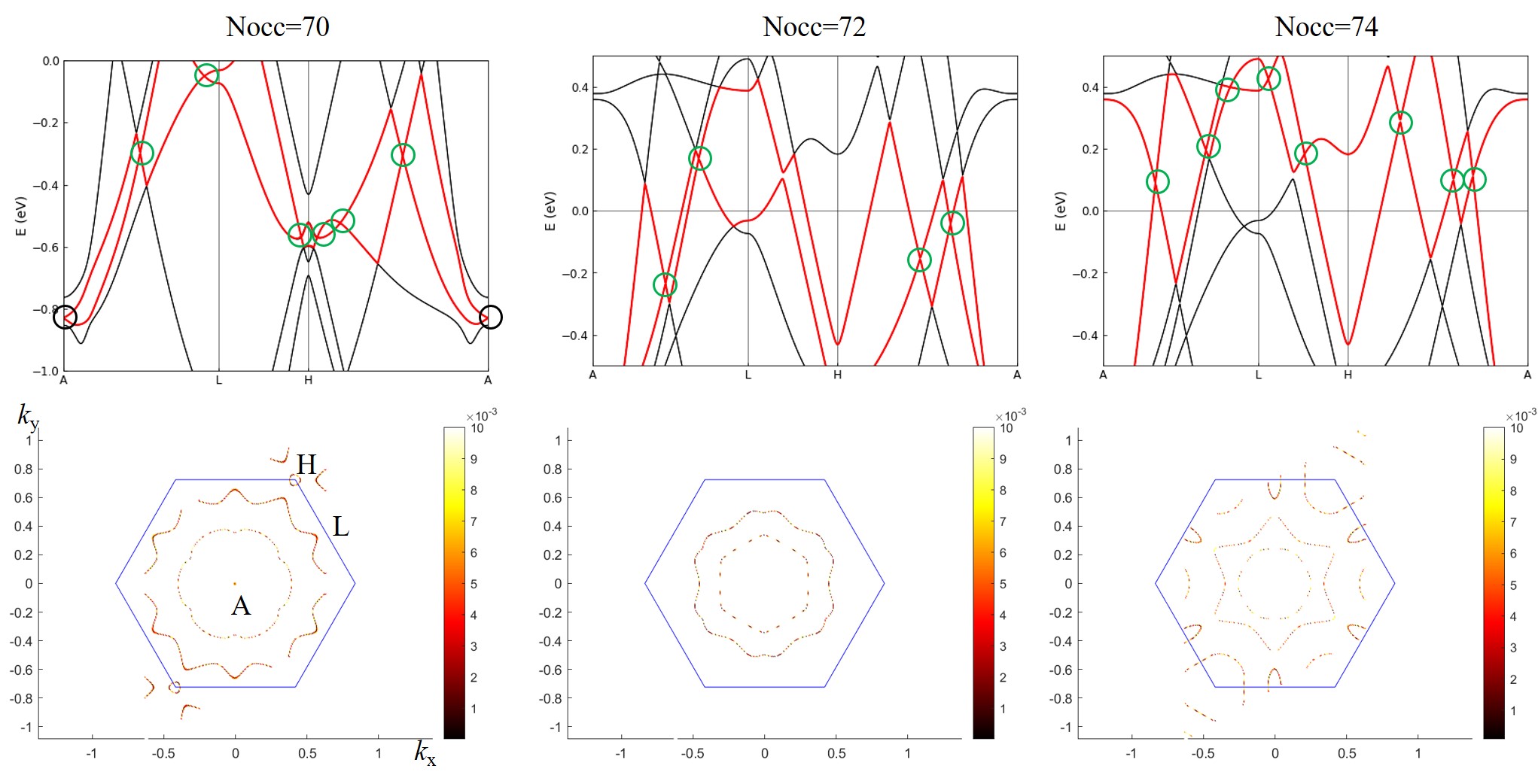}
\caption{~\label{figS5} Band structures and nodal loops of the pristine FeGe. The nodal loops are the crossing points between 70-71, 72-73 and 74-75 bands. The two pairs of bands between which we are searching the nodal points are colored red in the top panels. The nodal points are marked out by green circles. The bottoms panels are the direct gap (in unit of eV) distributions over the BZ.
}
\end{figure}

The band structures of the pristine FeGe are shown in Fig.~1(b) of main text. Due to the spatial inversion combined with time-reversal $+$ half lattice translation symmetry, the energy bands are all doubly degenerate though the system is antimagnetic. We focus on the three pairs of energy bands near the Fermi level, which are labelled band indices 70, 72 and 74. We then search the crossing points between each pairs of bands, which can be identified with the help of the direct gap (or topological gap) between 70-71, 72-73 bands and so on. We plot the results in Fig.~\ref{figS5}, where the touching points of the bands along the high-symmetry path are marked with black or green circles. Our calculations show that there exists Dirac points and/or Dirac nodal loops over the $k_{z}=\pi$ plane for all the three cases, which are protected by spatial inversion combined with time-reversal $+$ half lattice translation symmetry and the mirror symmetry with respect to the kagome plane. Specifically, when the highest number of occupied band is set to 70 (i.e., N$_{\rm{occ}}$=70), there is a Dirac point at the $A$ point. Additionally, there are two large nodal loops circling the $A$ point and one small triangular-like nodal loop circling the $H$ point. When N$_{\rm{occ}}$=72, there are multiple Dirac points and only one nodal loop circling the $A$ point. When N$_{\rm{occ}}$=74, there would be two nodal loops circling the $A$ point, one nodal loop circling the $H$ point and one nodal loop circling the $L$ point.

%\begin{figure}[!htbp]
%\includegraphics[width=0.8\textwidth]{FigS6.jpg}
%\caption{~\label{fig6} (a) Band structures of the pristine FeGe. (b) Phonon spectrum of FeGe, calculated using the frozen-phonon method.
%}
%\end{figure}

\subsection{DFT+$U$ band structures of pristine FeGe}
\begin{figure}[!htbp]
\includegraphics[width=0.7\textwidth]{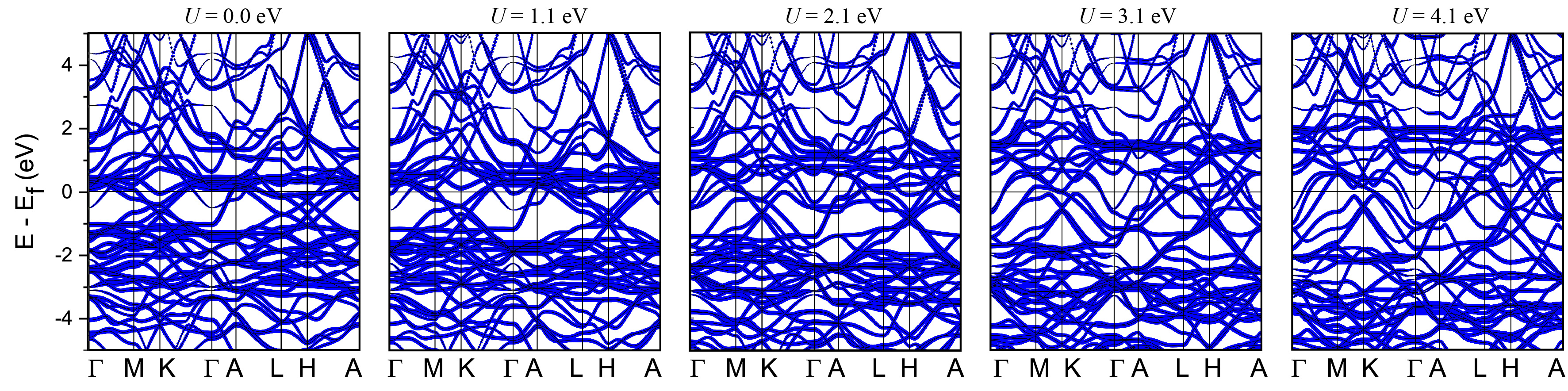}
\caption{~\label{figdftu} DFT+$U$ band structures of pristine FeGe, with $U=0, 1.1, 2.1, 3.1, 4.1\,$eV, respectively. The thickness of the blue lines represent the weight from Fe $3d$ orbitals.}
\end{figure}

In this subsection we show the DFT+$U$ band structures of pristine FeGe with different $U$ values, as shown in Fig.~\ref{figdftu}

%\begin{figure}[!htbp]
%\includegraphics[width=0.5\textwidth]{FigS8.jpg}
%\caption{~\label{fig6} Leading eigenvalues of the bare generalized susceptibility tensor  of pristine FeGe along a high-symmetry path over the first BZ.}
%\end{figure}

\section{\III. Generalized susceptibility tensors}

\subsection{Formalism}
We calculate the bare static generalized susceptibility tensor $\chi^0$ of pristine FeGe along a symmetry path within the first Brillouin zone, which is expressed as
\begin{equation}
\chi^{0}_{\mu\nu, \mu'\nu'}(\mathbf{q})=\int_{\rm BZ}\frac{\Omega\, d^3 \mathbf{k}}{(2\pi)^{3}}\sum_{m, n}\,\frac{f(E_{n,\mathbf{k}})-f(E_{m,\mathbf{k}+\mathbf{q}})}{E_{m,\mathbf{k}+\mathbf{q}}-E_{n,\mathbf{k}}}\,\psi_{\mu,n}^{*}(\mathbf{k})\psi_{\nu,m}(\mathbf{k}+\mathbf{q})\psi_{\mu',n}(\mathbf{k})\psi_{\nu',m}^{*}(\mathbf{k}+\mathbf{q})\ ,
\label{eq:chi0}
\end{equation}
where $\mu$, $\nu$ are composite indices referring to the sublattice, orbital and spin  degrees of freedom, $m$ and $n$ are the band indices, and $\Omega$ is the volume of the unit cell. $f(E_{n,\mathbf{k}})$ is the Fermi-Dirac distribution function at zero temperature, with  $E_{n,\mathbf{k}}$ denoting the band energy of the $n$th band at wavevector $\mathbf{k}$, and  $\psi_{\mu,n}(\mathbf{k})$ is the corresponding eigenfunction in the basis of the Fourier-transformed Wannier functions. The calculations are start from the DFT-based Wannier tight-binding Hamiltonian \cite{mostofi2008wannier90}. A linear tetrahedra interpolation method is applied for the integration of the $\mathbf{k}$ points over the first Brillouin zone \cite{interpolation-prb75} and a 30$\times$30$\times$18 $\mathbf{k}$ mesh is used. The summation over the bands are restricted to the 10 bands nearest to the Fermi energy. For the RPA susceptibility tensor, we use the Kanamori interactions as 
\begin{align}
H_{\textrm{K}}=&U\sum_{i,\alpha}\hat{n}_{i\alpha\uparrow}\hat{n}_{i\alpha\downarrow}+
U'\sum_{i,\alpha <\beta,\sigma,\sigma'}\hat{n}_{i\alpha\sigma}\hat{n}_{i\beta\sigma'}-J_{\textrm{H}}\sum_{i,\alpha < \beta,\sigma,\sigma'}\hat{c}^{\dagger}_{i\alpha\sigma}
\hat{c}^{\vphantom\dagger}_{i\alpha\sigma'}\hat{c}^{\dagger}_{i\beta\sigma'}\hat{c}^{\vphantom\dagger}_{i\beta\sigma}+J_{\textrm{P}}\sum_{i,\alpha < \beta,\sigma}
\hat{c}^{\dagger}_{i\alpha\sigma}\hat{c}^{\dagger}_{i\alpha-\sigma}\hat{c}^{\vphantom\dagger}_{i\beta\sigma}\hat{c}^{\vphantom\dagger}_{i\beta -\sigma}\;,
\label{eq:kanamori_R}
\end{align}
where $U$ and $U'$ are the intra-orbital and inter-orbital direct Coulomb interactions. 
$J_{\textrm{H}}$ and $J_{\textrm{P}}$ denote the Hunds' coupling 
and pair hoppings respectively, and $\hat{n}_{i\alpha\sigma}=\hat{c}^{\dagger}_{i\alpha\sigma}\hat{c}_{i\alpha\sigma}$ is the  density operator. We let $U'\!=\!U-J$ and $\!J_{\textrm{P}}\!= 0$ such that Eq.~(\ref{eq:kanamori_R}) has full rotational invariance \cite{georges2013strong}. We have considered effects of density-density interactions between neighboring Fe and Ge sites, 
\begin{equation}
H_{\textrm{Fe-Ge}}=V\,\sum_{\langle i  j \rangle}\sum_{\alpha\beta\sigma\sigma'}\,\hat{c}^{\dagger}_{i\alpha\sigma}\hat{c}^{\dagger}_{j\beta\sigma'}\hat{c}_{j\beta\sigma'}\hat{c}_{i\alpha\sigma}\;,
\label{eq:intersite}
\end{equation} 
with the interaction amplitude $V$. 
The effects of Coulomb interactions on the generalized susceptibility tensor are treated within random phase approximation (RPA), including effects of both direct and exchange interactions as described by the bubble-like and ladder-like Feynmann diagrams for the two-particle correlation function. All these Coulomb interaction terms (after Fourier transform) can be expressed in matrix form denoted by $\mathbbm{U}(\mathbf{q})$,  then the RPA generalized susceptibility tensor is expressed as
\begin{equation}
\hat{\mathbbm{\chi}}(\mathbf{q})=\hat{\mathbbm{\chi}^{0}(\mathbf{q})}\,\cdot\,(\,\mathbbm{1}+\mathbbm{U}(\mathbf{q})\cdot\hat{\mathbbm{\chi}}^{0}(\mathbf{q})\,)^{-1}
\label{eq:chi-RPA}
\end{equation}
where $\hat{\mathbbm{\chi}}(\mathbf{q})$ and $\hat{\mathbbm{\chi}^0}(\mathbf{q})$ refer to the RPA and bare susceptibility tensor (at wavevector $\mathbf{q}$) respectively. The ``$\cdot$" in Eq.~(\ref{eq:chi-RPA}) denotes a matrix product operation.

\subsection{More results}

\begin{figure}[!htbp]
\includegraphics[width=0.7\textwidth]{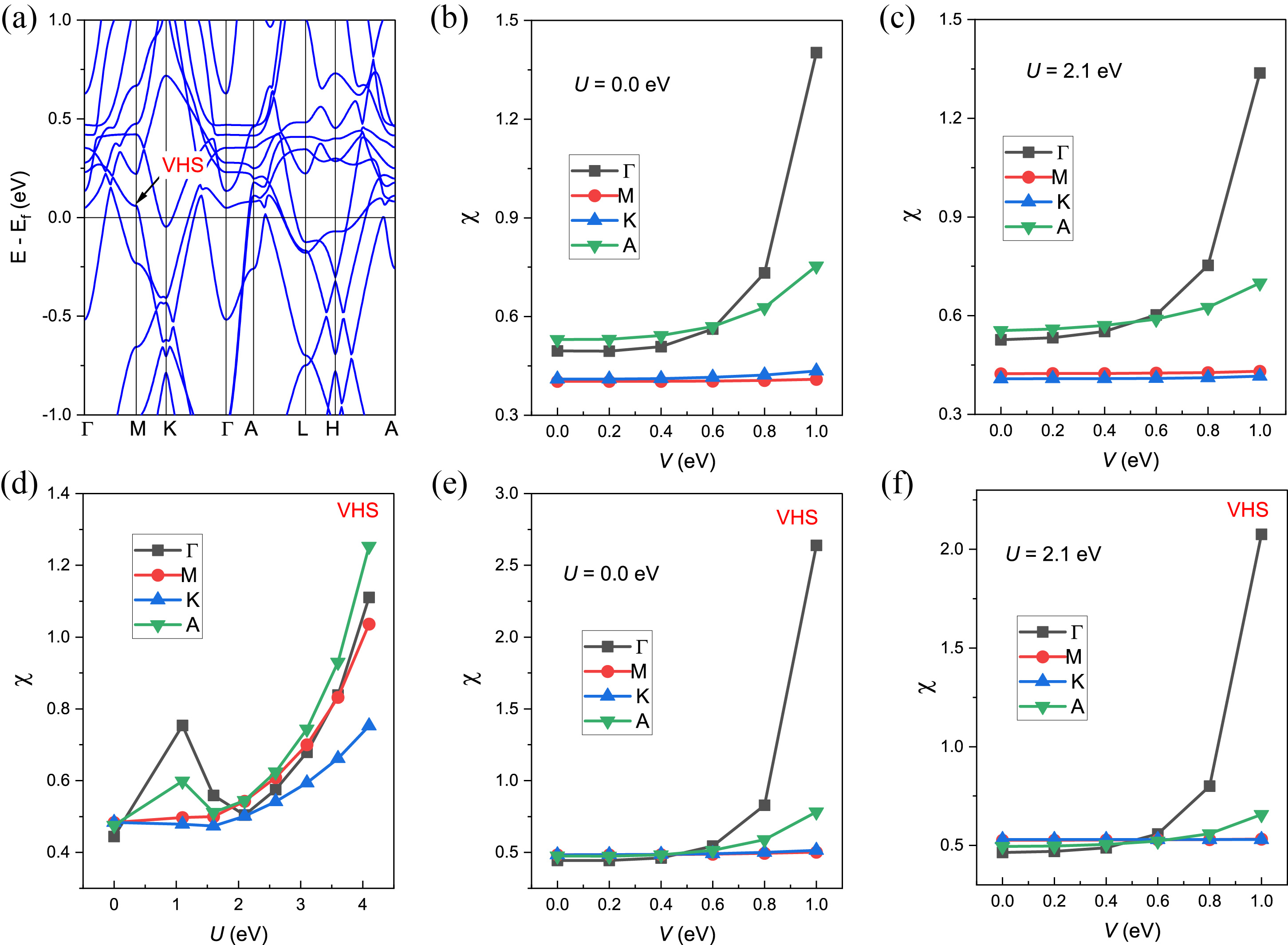}
\caption{~\label{figS12}  (a) Band structures of the pristine FeGe in antiferromagnetic state, with Hubbard $U=0$. The van Hove singularity (VHS) around  $M$ point is indicated by the black arrow. Leading eigenvalues of RPA susceptibility tensor at different high-symmetry points with different Fe-Se intersite interaction amplitudes $V$, the on-site $U$ is taken as 0.0\,eV in (b) and 2.1\,eV in (c). (d)-(e) Leading eigenvalues of the RPA susceptibility tensor, where the Fermi energy is set to the VHS.}
\end{figure}

\section{\IV. General tight-binding theory}
\begin{figure}[!htbp]
\includegraphics[width=0.5\textwidth]{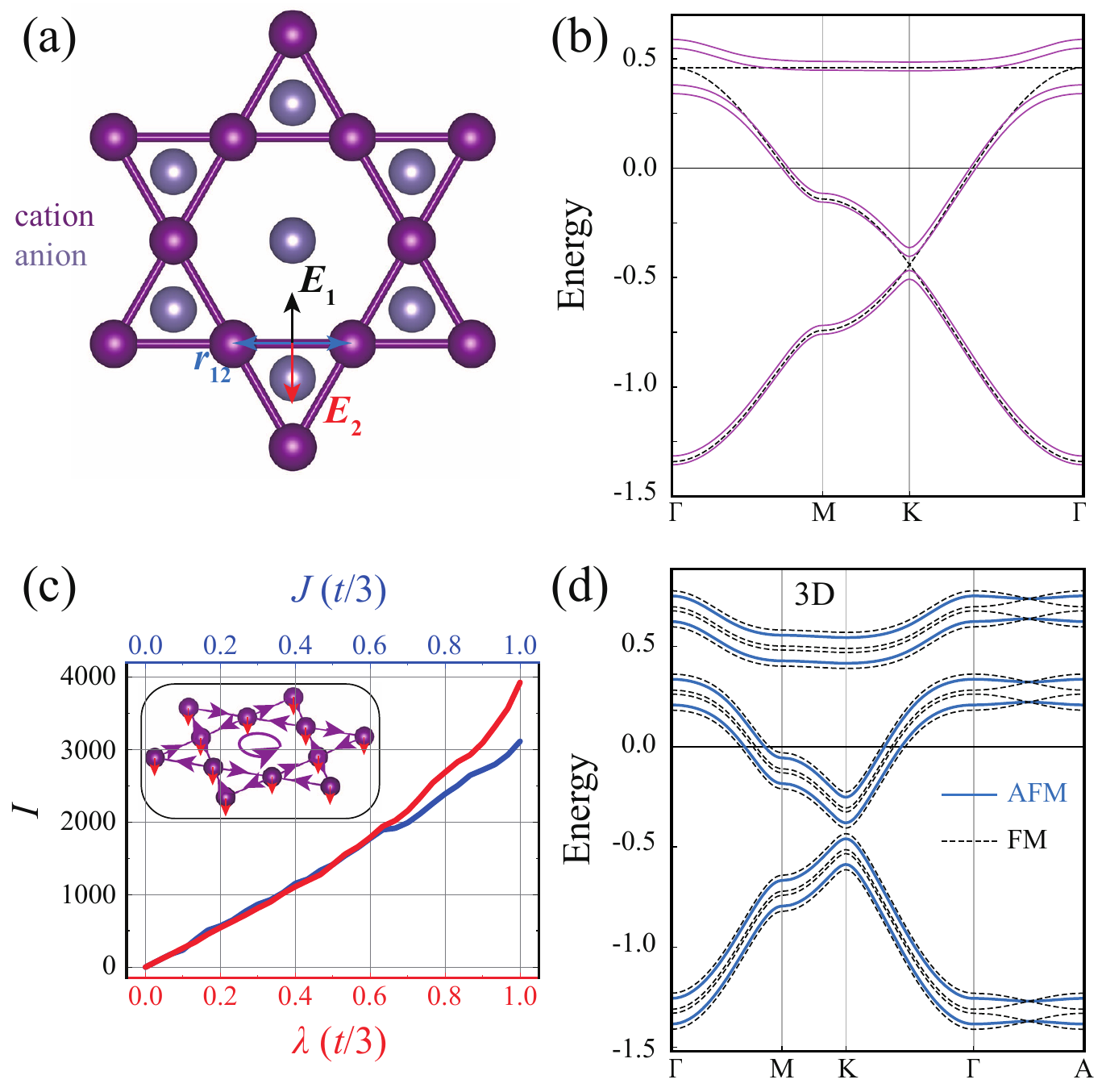}
\caption{~\label{figS1} Tight-binding theory of the chiral charge order in kagome magnets. (a) Lattice structure of the typical collinear kagome magnets with magnetic cations form the kagome lattice which are neighbored by the anions. %$d_1$ ($d_2$) denotes the cation-anion distance and 
The effective electric fields experienced by the neighboring cations towards the anions with position vector $\mathbf{r}_{ij}$ are denoted as $E_1$ and $E_2$. (b) Band structure of the kagome lattice without (black dash) and with magnetization and SOC (red solid), taken at $t=-0.3, \lambda=-0.1t, J=-0.2t/3$ and at half filling. (c) Dependence of the circulating currents $I$ on the SOC $\lambda$ (at $J=-0.2t/3$) and the Kondo coupling $J$ (at $\lambda=-0.1t$). The inset is the schematic of the inter-site current patterns, where the purple arrows indicate the directions of the current flow and the red arrows refer to the magnetization of the atoms. (d) Band structures of the 3D layered kagome lattice with interlayer ferromagnetically (FM, dash line) coupled or  antiferromagnetically (AFM, solid line). Taken at $t=-0.3, \lambda=-0.2t, J=-0.5t/3, t_{\perp}=0.2t/3$
and at half filling.}
\end{figure}

We start our discussion from the following simple tight-binding Hamiltonian for a monolayer ferromagnetic kagome lattice:
\begin{align}
\hat{H}_{\rm{mono}}=-t\sum_{\langle i j\rangle}\hat{c}^{\dagger}_{i}\hat{c}_{j}+i\lambda\sum_{\langle i\alpha,j\beta\rangle}(\mathbf{E}_{ij}\times\mathbf{r}_{ij})_{z}\cdot\sigma_{z}^{\alpha\beta}\hat{c}^{\dagger}_{i\alpha}\hat{c}_{j\beta}
+J\sum_{i\alpha\beta}\mathbf{S}_i\cdot\mathbf{\sigma}^{\alpha\beta}\hat{c}^{\dagger}_{i\alpha}\hat{c}_{i\beta}
\label{eq:Hmono}
\end{align}
where $\langle\rangle$ refers to the nearest neighbor hopping, $\hat{c}^{\dagger}_{i}\ (\hat{c}_{j})$ is the electron creation (annihilation) operator at site $i\ (j)$, $\alpha\beta$ is the spin index. %Similar Hamiltonian was considered in ref. \cite{Zhang2011} for the discussions of the quantum anomalous Hall effect in monolayer ferromagnetic kagome lattice. 
The first term is the usual first-neighbor hopping within the kagome triangles. The second term is the Kane-Mele type SOC term \cite{Kane2005} which is the generated by the internal in-plane electric field $\mathbf{E}_{ij}$ normal to the bond vector $\mathbf{r}_{ij}$ connecting the two sites $i$ and $j$ [see Fig.~\ref{figS1}(a)]. The third term is the effective magnetic exchange interactions. In general, more terms can be included into the model to be more realistic. Such as next-nearest-neighbor hoppings and SOC can be added to the describe the DT current pattern, and an effective charge order can also be added to simulate the SoD current patterns.

Generally, for the magnetic kagome compounds, the magnetic cations would form a kagome lattice, while the anions are typically located at the center of the hexagon and/or right on top of or  at the bottom of the center of the triangles (see Fig.~\ref{figS1}(a)). The $\mathbf{E}_{ij}$ experienced by the electrons hopping between nearby transition-metal cations thus is perpendicular to the bond vector $\mathbf{r}_{ij}$. Typically there are two types of cations in the three dimensional (3D)  kagome systems, one type of the cations are located at the center of the hexagons generating in-plane electric fields pointing towards hexagon center (denoted by $\mathbf{E}_1$ in Fig.~\ref{figS1}(a)), while the other ones are right on top of (or below) the triangle centers generating electric fields with the in-plane component pointing to the center of the triangles (denoted by $\mathbf{E}_2$). $\mathbf{E}_1$ and $\mathbf{E}_2$ would compete with each other, which determines the sign of the current loops thus the direction of orbital magnetization in kagome magnetic metals. 
The third term in Eq.~(\ref{eq:Hmono}) is the antiferromagnetic exchange coupling term with $J>0$ denoting the coupling strength, and $\mathbf{S}_i$ is the unit vector denoting the expectation value of the local magnetic moment at site $i$ which is FM within the kagome plane, and is coupled to the spin operator of itinerant electrons $\sum_{\alpha\beta}\hat{c}^{\dagger}_{i\alpha}\mathbf{\sigma}^{\alpha\beta}\hat{c}_{i\beta}$. Without SOC and the exchange coupling, the non-interacting band structures of the $H_{\rm{mono}}$ consist of a spin degenerate Dirac cone and a flat band as shown by the black lines in Fig.~\ref{figS1}(b). Including SOC and the exchange coupling term would gap out the Dirac point and lift the spin degeneracy as shown by the red lines in Fig.~\ref{figS1}(b). 

Circulating current loops within the kagome plane can be generated due to the interplay between ferromagnetism and SOC. Setting $t=-0.3$, $\lambda=-0.1t$, considering an out-of-plane magnetization pointing to negative $\mathbf{z}$ direction ($<S_z>=-1$) with $J=-0.067 t$, at half filling of the model, the calculated currents turn out to flow anticlockwise along the triangles, as shown in the inset of Fig.~\ref{figS1}(c). 
Moreover, treating the SOC ($\lambda$) and the exchange coupling ($J$) as a perturbation, the current amplitude turns out to be linearly dependent on both $\lambda$ and $J$ as shown in Fig.~\ref{figS1}(c).  
The ferromagnetic kagome monolayer (described by the Hamiltonian of Eq.~(\ref{eq:Hmono})) can be stacked along the $z$ direction including the inter-layer coupling (with amplitude $t_{\perp}$), which would yield a Hamiltonian describing the electronic structure of a 3D kagome magnet. In Fig.~\ref{figS1}(d) we present the band structures of 3D kagome magnetic metals (with $t_{\perp}=0.067 t$), where the dash and solid lines denote the interlayer ferromagnetic and AFM configurations, respectively. Such a model Hamiltonian and its variant can be applied to various ferromagnetic and antiferromagnetic 3D kagome magnetic metals such as FeGe, Co$_3$Sn$_2$S$_2$, Fe$_3$Sn, which are typical kagome compounds with collinear magnetic order. 

\subsection{Current patterns in some other kagome magnetic metals}
\label{sec:material}

\begin{figure}[!htb]
\includegraphics[width=0.6\textwidth]{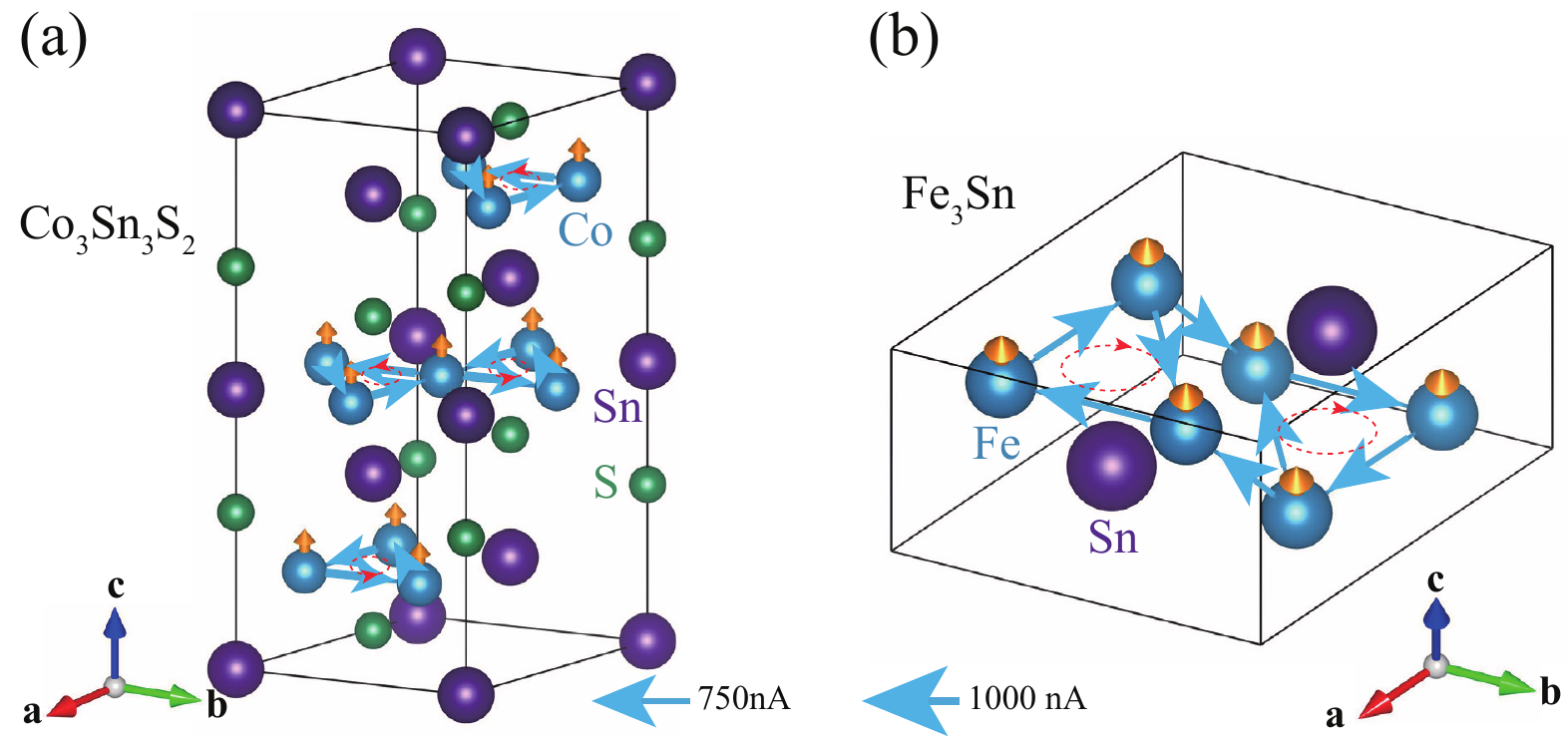}
\caption{~\label{figS2} Magnetic structures and inter-site current patterns of Co$_3$Sn$_2$S$_2$ and Fe$_3$Sn. The leading circulating current patterns are plotted by the arrows that connect the cation atoms, where the values of the inter-site currents are indicated by the thickness of the arrows. Note in \textbf{a} the arrows are shrunken by 50\% for better view.}
\end{figure}

We also consider Co$_3$Sn$_2$S$_2$, which  is a half-metallic ferromagnet.  The Curie temperature of Co$_3$Sn$_2$S$_2$ is 177\,K with the saturated magnetization 0.29~$\mu_{\rm B}$ per Co atom \cite{Yin2019}.  Its space group is $R3m$ (No. 166) with hexagonal lattice constant $a = 5.3\sim5.4\,$\AA, $c = 13.2\,$\AA \cite{Yin2019}. The kagome plane consists of the Co$_3$Sn layer with the Sn atom located at the center of the hexagon, while the S atoms are stacked on top of (and below) the center of the Co triangles [see Fig.~\ref{figS2}(a)]. A hexagonal Sn layer is further intercalated between two adjacent Co$_3$Sn layers. The system exhibits giant anomalous Hall effect (AHE) which is believed to be contributed from the Weyl fermions near the Fermi energy \cite{Liu2018,Wang2018}. Other than the AHE, STM experiments had observed negative flat-band orbital magnetism \cite{Yin2019}. 
 
\begin{figure}[tbp]
%\includegraphics[width=1.0\textwidth]{FigS2_NC.jpg}
%\caption{~\label{fig2} Calculated inter-site current patterns for FeGe, Co$_3$Sn$_2$S$_2$, Fe$_3$Sn and FeSn with SOC. \textbf{a} Inter-site current patterns for FeGe, taken at $\lambda/\lambda_0 = 1, U = 4.1$~eV, $J = U/5$. \textbf{b} Dependence of the circulating currents $I$ on the SOC $\lambda/\lambda_0$ and \textbf{c} on the local magnetization $M$. \textbf{d-f} Same as \textbf{a-c} but for Co$_3$Sn$_2$S$_2$, $U = 4.0$~eV, $J = U/5$. \textbf{g-i} For Fe$_3$Sn and \textbf{j-l} for FeSn, $U = 4.3$~eV, $J = U/5$.}

\includegraphics[width=0.73\textwidth]{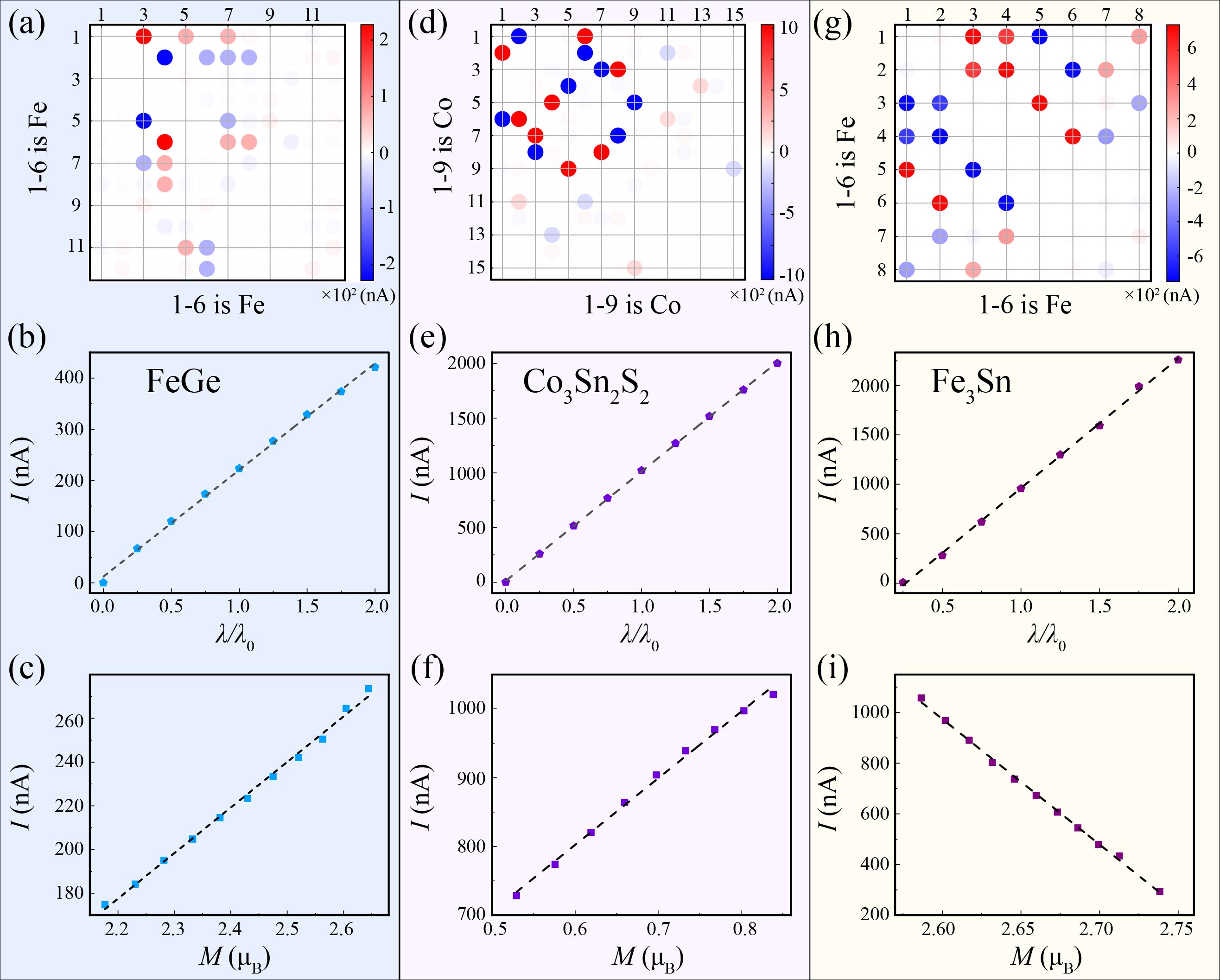}
\caption{~\label{figS3} Calculated inter-site current patterns for FeGe, Co$_3$Sn$_2$S$_2$, and Fe$_3$Sn with SOC. (a) Inter-site current patterns for FeGe, taken at $\lambda/\lambda_0 = 1, U = 4.1$~eV, $J = U/5$. (b) Dependence of the circulating currents $I$ on the SOC $\lambda/\lambda_0$ and (c) on the local magnetization $M$. (d-f) Same as (a-c) but for Co$_3$Sn$_2$S$_2$, $U = 4.0$~eV, $J = U/5$. (g-i) For Fe$_3$Sn.}
\end{figure}

In order to understand the negative orbital magnetism \cite{Yin2019}, we calculate the inter-site current patterns as shown in Fig.~\ref{figS2}(a) and Fig.~\ref{figS3}. As expected, the calculated inter-site currents flow along a loop within each ferromagnetic kagome layer generating an orbital flux through the kagome plane, which can be qualitatively described by the current pattern in Fig.~\ref{figS1}(c). The dramatic difference of  electric negativities between the Co cation and the S anion would give rise to a strong in-plane electric field and a large Kane-Mele type SOC amplitude $\lambda$. Because the inter-site current amplitude is linearly dependent on $\lambda$ and $J$ (Fig.~\ref{figS3}), the large Kane-Mele SOC generates an inter-site current as large as ~1000\,nA and an orbital magnetization on the order of $0.1\,\mu_{B}$ per cell that is anti-parallel to the spin magnetization. 
This explains the origin of the giant negative orbital magnetism observed in experiments \cite{Yin2019}.

%\subsection{Fe$_3$Sn}
%\subsubsection{Fe$_3$Sn}
Our theory can also be applied to Fe$_3$Sn, whose primitive cell consists of two Fe$_3$Sn kagome layers with short interlayer distance (thus strong interlayer couplings), with an hexagonal lattice constants $a\approx5.46\,$\AA\ and $c\approx 4.36\,$\AA,  as shown in Fig.~\ref{fig4}b. The system is ferromagnetic with a Curie temperature $T_c=743\,$K \cite{Trumpy-PRB-1970}. By virtue of the strong short interlayer distance and strong interlayer coupling, the calculated current pattern in Fig.~\ref{figS2}b involves both intra-layer and interlayer current loops with the current amplitude as large as $\sim720\,$nA, forming a intercalated network flowing through the 3D kagome lattice. Moreover, different from Co$_3$Sn$_2$S$_2$, in which the orbital magnetization generated by the circulating current is anti-parallel to the spin magnetization, here they are parallel to each other in Fe$_3$Sn (Fig.~\ref{figS3}).

\subsection{Tight-binding model for $2\times2$ supercell structure}

\begin{figure}[bth!]
\includegraphics[width=0.7\textwidth]{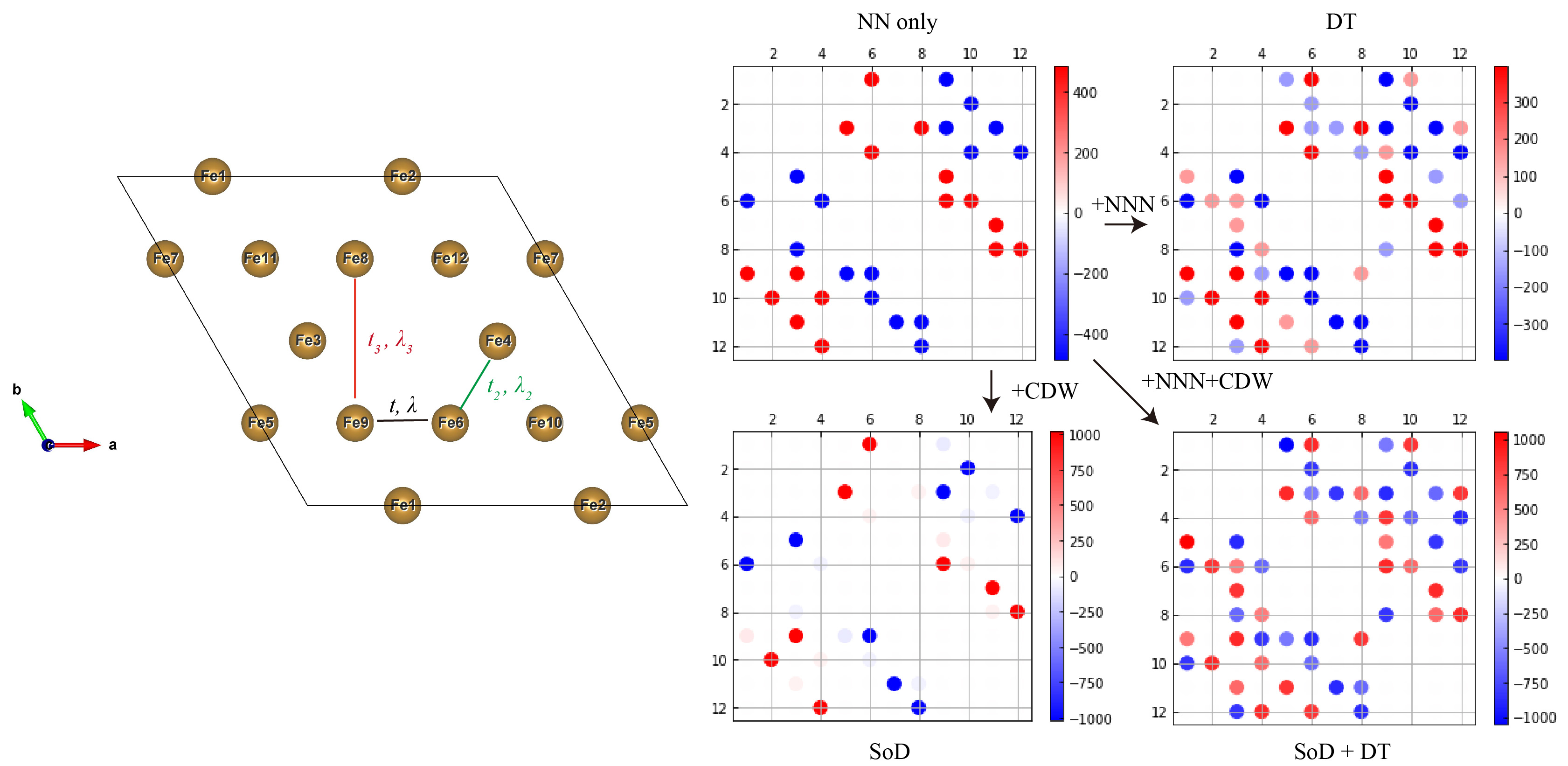}
\caption{~\label{fig11} 
Tight-binding models of the $2\times2$ supercell structure. The hoppings parameters for the nearest-neighbor (NN, $t, \lambda$), next-nearest-neighbor (NNN, $t_2, \lambda_2$) magnetic atoms and those modified by the CDW distortions ($t_3, \lambda_3$) are illustrated in the lattice structure in the left panel. Right panel, current patterns of the $2\times2$ superceel structures with additional NNN interactions and/or SoD CDW order. The parameters are $t=-0.3,\ t_{2}=t, t_{3}=0, \lambda=-0.1t,\  \lambda_{2}=\lambda, \lambda_{3}=0\  J=-0.1t,\ {\rm Nocc=}12/24$ for the top-left inset, $t=-0.3,\ t_{2}=t, t_{3}=t/6, \lambda=-0.1t,\  \lambda_{2}=\lambda, \lambda_{3}=5/3\lambda$ for the top-right inset, $t=-0.3,\ t_{2}=t/1.3, t_{3}=0, \lambda=-0.1t,\  \lambda_{2}=\lambda/1.3, \lambda_{3}=0$ for the bottom-left and $t=-0.3,\ t_{2}=t/1.3, t_{3}=14t/15, \lambda=-0.1t,\  \lambda_{2}=\lambda/1.3, \lambda_{3}=14\lambda/15$ for the bottom-right}
\end{figure}

In the main text, we mentioned that more terms can be added to Eq.~(\ref{eq:Hmono}) to describe a more realistic structure such as the $2\times2\times1$ CDW phase of FeGe. Here we make some extensions on Eq.~(\ref{eq:Hmono}).
Start from the $2\times2$ supercell structure with only the nearest-neighbor (NN) interactions, we fisrt added the next-nearest-neighbor (NNN) hoppings and SOC. Usually, the NNN interactions are small compared with the NN interactions and in generally play a negligible role in the current patterns. However, in some case such as the CDW phase of FeGe, the two become comparable, then the DT current patterns induced by the NNN interactions would also become important. Inclucing the NNN interactions to Eq.~(\ref{eq:Hmono}) is direct, and the resulted current pattern is shown in Fig.~\ref{fig11}
We then consider the CDW order, which can be included to Eq.~(\ref{eq:Hmono}) phenomenally as distortions to the NN and NNN interactions. Here we only consider its effects on the NN interactions, which modify some $t, \lambda$ to $t_2, \lambda_2$. Without NNN interactions, the resulted SoD current pattern can be clearly found in Fig.~\ref{fig11}when $t_2, \lambda_2$ is a little bit smaller than $t, \lambda$.
Finally, we incorporate all the ingradients into together, the resulted current patterns contained both DT and SoD patterns, which can well reproduced the results in $2\times2\times1$ CDW phase of FeGe. By now, the simple extension to the tight-binding model done well, more delicate modifications to Eq.~(\ref{eq:Hmono}) should be made when trying to apply to more realistic cases.

\newpage 

\bibliography{TCFS}

\begin{thebibliography}{90}
\expandafter\ifx\csname natexlab\endcsname\relax\def\natexlab#1{#1}\fi
\expandafter\ifx\csname bibnamefont\endcsname\relax
  \def\bibnamefont#1{#1}\fi
\expandafter\ifx\csname bibfnamefont\endcsname\relax
  \def\bibfnamefont#1{#1}\fi
\expandafter\ifx\csname citenamefont\endcsname\relax
  \def\citenamefont#1{#1}\fi
\expandafter\ifx\csname url\endcsname\relax
  \def\url#1{\texttt{#1}}\fi
\expandafter\ifx\csname urlprefix\endcsname\relax\def\urlprefix{URL }\fi
\providecommand{\bibinfo}[2]{#2}
\providecommand{\eprint}[2][]{\url{#2}}

\bibitem[{\citenamefont{Bergman et~al.}(2008)\citenamefont{Bergman, Wu, and
  Balents}}]{balents-prb08}
\bibinfo{author}{\bibfnamefont{D.~L.} \bibnamefont{Bergman}},
  \bibinfo{author}{\bibfnamefont{C.}~\bibnamefont{Wu}}, \bibnamefont{and}
  \bibinfo{author}{\bibfnamefont{L.}~\bibnamefont{Balents}},
  \bibinfo{journal}{Phys. Rev. B} \textbf{\bibinfo{volume}{78}},
  \bibinfo{pages}{125104} (\bibinfo{year}{2008}).

\bibitem[{\citenamefont{Zhou et~al.}(2017)\citenamefont{Zhou, Kanoda, and
  Ng}}]{Zhou2017}
\bibinfo{author}{\bibfnamefont{Y.}~\bibnamefont{Zhou}},
  \bibinfo{author}{\bibfnamefont{K.}~\bibnamefont{Kanoda}}, \bibnamefont{and}
  \bibinfo{author}{\bibfnamefont{T.-K.} \bibnamefont{Ng}},
  \bibinfo{journal}{Reviews of Modern Physics} \textbf{\bibinfo{volume}{89}},
  \bibinfo{pages}{025003} (\bibinfo{year}{2017}), ISSN
  \bibinfo{issn}{0034-6861}.

\bibitem[{\citenamefont{Neupert et~al.}(2022)\citenamefont{Neupert, Denner,
  Yin, Thomale, and Hasan}}]{Neupert2022}
\bibinfo{author}{\bibfnamefont{T.}~\bibnamefont{Neupert}},
  \bibinfo{author}{\bibfnamefont{M.~M.} \bibnamefont{Denner}},
  \bibinfo{author}{\bibfnamefont{J.-X.} \bibnamefont{Yin}},
  \bibinfo{author}{\bibfnamefont{R.}~\bibnamefont{Thomale}}, \bibnamefont{and}
  \bibinfo{author}{\bibfnamefont{M.~Z.} \bibnamefont{Hasan}},
  \bibinfo{journal}{Nature Physics} \textbf{\bibinfo{volume}{18}},
  \bibinfo{pages}{137} (\bibinfo{year}{2022}), ISSN \bibinfo{issn}{1745-2473}.

\bibitem[{\citenamefont{Yin et~al.}(2022)\citenamefont{Yin, Lian, and
  Hasan}}]{yin-review-nature22}
\bibinfo{author}{\bibfnamefont{J.-X.} \bibnamefont{Yin}},
  \bibinfo{author}{\bibfnamefont{B.}~\bibnamefont{Lian}}, \bibnamefont{and}
  \bibinfo{author}{\bibfnamefont{M.~Z.} \bibnamefont{Hasan}},
  \bibinfo{journal}{Nature} \textbf{\bibinfo{volume}{612}},
  \bibinfo{pages}{647} (\bibinfo{year}{2022}), ISSN \bibinfo{issn}{1476-4687}.

\bibitem[{\citenamefont{Tang et~al.}(2011)\citenamefont{Tang, Mei, and
  Wen}}]{wen-prl11}
\bibinfo{author}{\bibfnamefont{E.}~\bibnamefont{Tang}},
  \bibinfo{author}{\bibfnamefont{J.-W.} \bibnamefont{Mei}}, \bibnamefont{and}
  \bibinfo{author}{\bibfnamefont{X.-G.} \bibnamefont{Wen}},
  \bibinfo{journal}{Phys. Rev. Lett.} \textbf{\bibinfo{volume}{106}},
  \bibinfo{pages}{236802} (\bibinfo{year}{2011}).

\bibitem[{\citenamefont{Kiesel et~al.}(2013{\natexlab{a}})\citenamefont{Kiesel,
  Platt, and Thomale}}]{thomale-kagome-prl13}
\bibinfo{author}{\bibfnamefont{M.~L.} \bibnamefont{Kiesel}},
  \bibinfo{author}{\bibfnamefont{C.}~\bibnamefont{Platt}}, \bibnamefont{and}
  \bibinfo{author}{\bibfnamefont{R.}~\bibnamefont{Thomale}},
  \bibinfo{journal}{Phys. Rev. Lett.} \textbf{\bibinfo{volume}{110}},
  \bibinfo{pages}{126405} (\bibinfo{year}{2013}{\natexlab{a}}).

\bibitem[{\citenamefont{Wang et~al.}(2013)\citenamefont{Wang, Li, Xiang, and
  Wang}}]{wang-kagome-prb13}
\bibinfo{author}{\bibfnamefont{W.-S.} \bibnamefont{Wang}},
  \bibinfo{author}{\bibfnamefont{Z.-Z.} \bibnamefont{Li}},
  \bibinfo{author}{\bibfnamefont{Y.-Y.} \bibnamefont{Xiang}}, \bibnamefont{and}
  \bibinfo{author}{\bibfnamefont{Q.-H.} \bibnamefont{Wang}},
  \bibinfo{journal}{Phys. Rev. B} \textbf{\bibinfo{volume}{87}},
  \bibinfo{pages}{115135} (\bibinfo{year}{2013}).

\bibitem[{\citenamefont{Kiesel and Thomale}(2012)}]{thomale-kagome-prb12}
\bibinfo{author}{\bibfnamefont{M.~L.} \bibnamefont{Kiesel}} \bibnamefont{and}
  \bibinfo{author}{\bibfnamefont{R.}~\bibnamefont{Thomale}},
  \bibinfo{journal}{Phys. Rev. B} \textbf{\bibinfo{volume}{86}},
  \bibinfo{pages}{121105} (\bibinfo{year}{2012}).

\bibitem[{\citenamefont{Yu and Li}(2012)}]{li-kagome-prb12}
\bibinfo{author}{\bibfnamefont{S.-L.} \bibnamefont{Yu}} \bibnamefont{and}
  \bibinfo{author}{\bibfnamefont{J.-X.} \bibnamefont{Li}},
  \bibinfo{journal}{Phys. Rev. B} \textbf{\bibinfo{volume}{85}},
  \bibinfo{pages}{144402} (\bibinfo{year}{2012}).

\bibitem[{\citenamefont{Yin et~al.}(2018{\natexlab{a}})\citenamefont{Yin,
  Zhang, Li, Jiang, Chang, Zhang, Lian, Xiang, Belopolski, Zheng
  et~al.}}]{yin2018giant}
\bibinfo{author}{\bibfnamefont{J.-X.} \bibnamefont{Yin}},
  \bibinfo{author}{\bibfnamefont{S.~S.} \bibnamefont{Zhang}},
  \bibinfo{author}{\bibfnamefont{H.}~\bibnamefont{Li}},
  \bibinfo{author}{\bibfnamefont{K.}~\bibnamefont{Jiang}},
  \bibinfo{author}{\bibfnamefont{G.}~\bibnamefont{Chang}},
  \bibinfo{author}{\bibfnamefont{B.}~\bibnamefont{Zhang}},
  \bibinfo{author}{\bibfnamefont{B.}~\bibnamefont{Lian}},
  \bibinfo{author}{\bibfnamefont{C.}~\bibnamefont{Xiang}},
  \bibinfo{author}{\bibfnamefont{I.}~\bibnamefont{Belopolski}},
  \bibinfo{author}{\bibfnamefont{H.}~\bibnamefont{Zheng}},
  \bibnamefont{et~al.}, \bibinfo{journal}{Nature}
  \textbf{\bibinfo{volume}{562}}, \bibinfo{pages}{91}
  (\bibinfo{year}{2018}{\natexlab{a}}).

\bibitem[{\citenamefont{Ye et~al.}(2018{\natexlab{a}})\citenamefont{Ye, Kang,
  Liu, Von~Cube, Wicker, Suzuki, Jozwiak, Bostwick, Rotenberg, Bell
  et~al.}}]{ye2018massive}
\bibinfo{author}{\bibfnamefont{L.}~\bibnamefont{Ye}},
  \bibinfo{author}{\bibfnamefont{M.}~\bibnamefont{Kang}},
  \bibinfo{author}{\bibfnamefont{J.}~\bibnamefont{Liu}},
  \bibinfo{author}{\bibfnamefont{F.}~\bibnamefont{Von~Cube}},
  \bibinfo{author}{\bibfnamefont{C.~R.} \bibnamefont{Wicker}},
  \bibinfo{author}{\bibfnamefont{T.}~\bibnamefont{Suzuki}},
  \bibinfo{author}{\bibfnamefont{C.}~\bibnamefont{Jozwiak}},
  \bibinfo{author}{\bibfnamefont{A.}~\bibnamefont{Bostwick}},
  \bibinfo{author}{\bibfnamefont{E.}~\bibnamefont{Rotenberg}},
  \bibinfo{author}{\bibfnamefont{D.~C.} \bibnamefont{Bell}},
  \bibnamefont{et~al.}, \bibinfo{journal}{Nature}
  \textbf{\bibinfo{volume}{555}}, \bibinfo{pages}{638}
  (\bibinfo{year}{2018}{\natexlab{a}}).

\bibitem[{\citenamefont{Liu et~al.}(2018{\natexlab{a}})\citenamefont{Liu, Sun,
  Kumar, Muechler, Sun, Jiao, Yang, Liu, Liang, Xu et~al.}}]{liu2018giant}
\bibinfo{author}{\bibfnamefont{E.}~\bibnamefont{Liu}},
  \bibinfo{author}{\bibfnamefont{Y.}~\bibnamefont{Sun}},
  \bibinfo{author}{\bibfnamefont{N.}~\bibnamefont{Kumar}},
  \bibinfo{author}{\bibfnamefont{L.}~\bibnamefont{Muechler}},
  \bibinfo{author}{\bibfnamefont{A.}~\bibnamefont{Sun}},
  \bibinfo{author}{\bibfnamefont{L.}~\bibnamefont{Jiao}},
  \bibinfo{author}{\bibfnamefont{S.-Y.} \bibnamefont{Yang}},
  \bibinfo{author}{\bibfnamefont{D.}~\bibnamefont{Liu}},
  \bibinfo{author}{\bibfnamefont{A.}~\bibnamefont{Liang}},
  \bibinfo{author}{\bibfnamefont{Q.}~\bibnamefont{Xu}}, \bibnamefont{et~al.},
  \bibinfo{journal}{Nature physics} \textbf{\bibinfo{volume}{14}},
  \bibinfo{pages}{1125} (\bibinfo{year}{2018}{\natexlab{a}}).

\bibitem[{\citenamefont{Yin et~al.}(2020)\citenamefont{Yin, Ma, Cochran, Xu,
  Zhang, Tien, Shumiya, Cheng, Jiang, Lian et~al.}}]{yin2020quantum}
\bibinfo{author}{\bibfnamefont{J.-X.} \bibnamefont{Yin}},
  \bibinfo{author}{\bibfnamefont{W.}~\bibnamefont{Ma}},
  \bibinfo{author}{\bibfnamefont{T.~A.} \bibnamefont{Cochran}},
  \bibinfo{author}{\bibfnamefont{X.}~\bibnamefont{Xu}},
  \bibinfo{author}{\bibfnamefont{S.~S.} \bibnamefont{Zhang}},
  \bibinfo{author}{\bibfnamefont{H.-J.} \bibnamefont{Tien}},
  \bibinfo{author}{\bibfnamefont{N.}~\bibnamefont{Shumiya}},
  \bibinfo{author}{\bibfnamefont{G.}~\bibnamefont{Cheng}},
  \bibinfo{author}{\bibfnamefont{K.}~\bibnamefont{Jiang}},
  \bibinfo{author}{\bibfnamefont{B.}~\bibnamefont{Lian}}, \bibnamefont{et~al.},
  \bibinfo{journal}{Nature} \textbf{\bibinfo{volume}{583}},
  \bibinfo{pages}{533} (\bibinfo{year}{2020}).

\bibitem[{\citenamefont{Ko et~al.}(2009)\citenamefont{Ko, Lee, and
  Wen}}]{ko2009doped}
\bibinfo{author}{\bibfnamefont{W.-H.} \bibnamefont{Ko}},
  \bibinfo{author}{\bibfnamefont{P.~A.} \bibnamefont{Lee}}, \bibnamefont{and}
  \bibinfo{author}{\bibfnamefont{X.-G.} \bibnamefont{Wen}},
  \bibinfo{journal}{Physical Review B} \textbf{\bibinfo{volume}{79}},
  \bibinfo{pages}{214502} (\bibinfo{year}{2009}).

\bibitem[{\citenamefont{Balents}(2010)}]{balents2010spin}
\bibinfo{author}{\bibfnamefont{L.}~\bibnamefont{Balents}},
  \bibinfo{journal}{Nature} \textbf{\bibinfo{volume}{464}},
  \bibinfo{pages}{199} (\bibinfo{year}{2010}).

\bibitem[{\citenamefont{Kiesel et~al.}(2013{\natexlab{b}})\citenamefont{Kiesel,
  Platt, and Thomale}}]{kiesel2013unconventional}
\bibinfo{author}{\bibfnamefont{M.~L.} \bibnamefont{Kiesel}},
  \bibinfo{author}{\bibfnamefont{C.}~\bibnamefont{Platt}}, \bibnamefont{and}
  \bibinfo{author}{\bibfnamefont{R.}~\bibnamefont{Thomale}},
  \bibinfo{journal}{Physical review letters} \textbf{\bibinfo{volume}{110}},
  \bibinfo{pages}{126405} (\bibinfo{year}{2013}{\natexlab{b}}).

\bibitem[{\citenamefont{Yin et~al.}(2018{\natexlab{b}})\citenamefont{Yin,
  Zhang, Li, Jiang, Chang, Zhang, Lian, Xiang, Belopolski, Zheng
  et~al.}}]{Yin2018}
\bibinfo{author}{\bibfnamefont{J.-X.} \bibnamefont{Yin}},
  \bibinfo{author}{\bibfnamefont{S.~S.} \bibnamefont{Zhang}},
  \bibinfo{author}{\bibfnamefont{H.}~\bibnamefont{Li}},
  \bibinfo{author}{\bibfnamefont{K.}~\bibnamefont{Jiang}},
  \bibinfo{author}{\bibfnamefont{G.}~\bibnamefont{Chang}},
  \bibinfo{author}{\bibfnamefont{B.}~\bibnamefont{Zhang}},
  \bibinfo{author}{\bibfnamefont{B.}~\bibnamefont{Lian}},
  \bibinfo{author}{\bibfnamefont{C.}~\bibnamefont{Xiang}},
  \bibinfo{author}{\bibfnamefont{I.}~\bibnamefont{Belopolski}},
  \bibinfo{author}{\bibfnamefont{H.}~\bibnamefont{Zheng}},
  \bibnamefont{et~al.}, \bibinfo{journal}{Nature}
  \textbf{\bibinfo{volume}{562}}, \bibinfo{pages}{91}
  (\bibinfo{year}{2018}{\natexlab{b}}), ISSN \bibinfo{issn}{0028-0836}.

\bibitem[{\citenamefont{Ye et~al.}(2018{\natexlab{b}})\citenamefont{Ye, Kang,
  Liu, von Cube, Wicker, Suzuki, Jozwiak, Bostwick, Rotenberg, Bell
  et~al.}}]{Ye2018}
\bibinfo{author}{\bibfnamefont{L.}~\bibnamefont{Ye}},
  \bibinfo{author}{\bibfnamefont{M.}~\bibnamefont{Kang}},
  \bibinfo{author}{\bibfnamefont{J.}~\bibnamefont{Liu}},
  \bibinfo{author}{\bibfnamefont{F.}~\bibnamefont{von Cube}},
  \bibinfo{author}{\bibfnamefont{C.~R.} \bibnamefont{Wicker}},
  \bibinfo{author}{\bibfnamefont{T.}~\bibnamefont{Suzuki}},
  \bibinfo{author}{\bibfnamefont{C.}~\bibnamefont{Jozwiak}},
  \bibinfo{author}{\bibfnamefont{A.}~\bibnamefont{Bostwick}},
  \bibinfo{author}{\bibfnamefont{E.}~\bibnamefont{Rotenberg}},
  \bibinfo{author}{\bibfnamefont{D.~C.} \bibnamefont{Bell}},
  \bibnamefont{et~al.}, \bibinfo{journal}{Nature}
  \textbf{\bibinfo{volume}{555}}, \bibinfo{pages}{638}
  (\bibinfo{year}{2018}{\natexlab{b}}), ISSN \bibinfo{issn}{0028-0836}.

\bibitem[{\citenamefont{Xu et~al.}(2015)\citenamefont{Xu, Lian, and
  Zhang}}]{Xu2015}
\bibinfo{author}{\bibfnamefont{G.}~\bibnamefont{Xu}},
  \bibinfo{author}{\bibfnamefont{B.}~\bibnamefont{Lian}}, \bibnamefont{and}
  \bibinfo{author}{\bibfnamefont{S.-C.} \bibnamefont{Zhang}},
  \bibinfo{journal}{Physical Review Letters} \textbf{\bibinfo{volume}{115}},
  \bibinfo{pages}{186802} (\bibinfo{year}{2015}), ISSN
  \bibinfo{issn}{0031-9007},
  \urlprefix\url{https://link.aps.org/doi/10.1103/PhysRevLett.115.186802}.

\bibitem[{\citenamefont{Wu et~al.}(2021{\natexlab{a}})\citenamefont{Wu,
  Schwemmer, M{\"u}ller, Consiglio, Sangiovanni, Di~Sante, Iqbal, Hanke,
  Schnyder, Denner et~al.}}]{wu2021nature}
\bibinfo{author}{\bibfnamefont{X.}~\bibnamefont{Wu}},
  \bibinfo{author}{\bibfnamefont{T.}~\bibnamefont{Schwemmer}},
  \bibinfo{author}{\bibfnamefont{T.}~\bibnamefont{M{\"u}ller}},
  \bibinfo{author}{\bibfnamefont{A.}~\bibnamefont{Consiglio}},
  \bibinfo{author}{\bibfnamefont{G.}~\bibnamefont{Sangiovanni}},
  \bibinfo{author}{\bibfnamefont{D.}~\bibnamefont{Di~Sante}},
  \bibinfo{author}{\bibfnamefont{Y.}~\bibnamefont{Iqbal}},
  \bibinfo{author}{\bibfnamefont{W.}~\bibnamefont{Hanke}},
  \bibinfo{author}{\bibfnamefont{A.~P.} \bibnamefont{Schnyder}},
  \bibinfo{author}{\bibfnamefont{M.~M.} \bibnamefont{Denner}},
  \bibnamefont{et~al.}, \bibinfo{journal}{Physical review letters}
  \textbf{\bibinfo{volume}{127}}, \bibinfo{pages}{177001}
  (\bibinfo{year}{2021}{\natexlab{a}}).

\bibitem[{\citenamefont{Feng et~al.}(2021)\citenamefont{Feng, Jiang, Wang, and
  Hu}}]{feng2021chiral}
\bibinfo{author}{\bibfnamefont{X.}~\bibnamefont{Feng}},
  \bibinfo{author}{\bibfnamefont{K.}~\bibnamefont{Jiang}},
  \bibinfo{author}{\bibfnamefont{Z.}~\bibnamefont{Wang}}, \bibnamefont{and}
  \bibinfo{author}{\bibfnamefont{J.}~\bibnamefont{Hu}},
  \bibinfo{journal}{Science Bulletin}  (\bibinfo{year}{2021}).

\bibitem[{\citenamefont{Gu et~al.}(2021)\citenamefont{Gu, Zhang, Feng, Jiang,
  and Hu}}]{gu2021gapless}
\bibinfo{author}{\bibfnamefont{Y.}~\bibnamefont{Gu}},
  \bibinfo{author}{\bibfnamefont{Y.}~\bibnamefont{Zhang}},
  \bibinfo{author}{\bibfnamefont{X.}~\bibnamefont{Feng}},
  \bibinfo{author}{\bibfnamefont{K.}~\bibnamefont{Jiang}}, \bibnamefont{and}
  \bibinfo{author}{\bibfnamefont{J.}~\bibnamefont{Hu}}, \bibinfo{journal}{arXiv
  preprint arXiv:2108.04703}  (\bibinfo{year}{2021}).

\bibitem[{\citenamefont{Park et~al.}(2021)\citenamefont{Park, Ye, and
  Balents}}]{balents-instability-prb21}
\bibinfo{author}{\bibfnamefont{T.}~\bibnamefont{Park}},
  \bibinfo{author}{\bibfnamefont{M.}~\bibnamefont{Ye}}, \bibnamefont{and}
  \bibinfo{author}{\bibfnamefont{L.}~\bibnamefont{Balents}},
  \bibinfo{journal}{Phys. Rev. B} \textbf{\bibinfo{volume}{104}},
  \bibinfo{pages}{035142} (\bibinfo{year}{2021}).

\bibitem[{\citenamefont{Ortiz et~al.}(2020)\citenamefont{Ortiz, Teicher, Hu,
  Zuo, Sarte, Schueller, Abeykoon, Krogstad, Rosenkranz, Osborn
  et~al.}}]{Ortiz2020}
\bibinfo{author}{\bibfnamefont{B.~R.} \bibnamefont{Ortiz}},
  \bibinfo{author}{\bibfnamefont{S.~M.} \bibnamefont{Teicher}},
  \bibinfo{author}{\bibfnamefont{Y.}~\bibnamefont{Hu}},
  \bibinfo{author}{\bibfnamefont{J.~L.} \bibnamefont{Zuo}},
  \bibinfo{author}{\bibfnamefont{P.~M.} \bibnamefont{Sarte}},
  \bibinfo{author}{\bibfnamefont{E.~C.} \bibnamefont{Schueller}},
  \bibinfo{author}{\bibfnamefont{A.~M.} \bibnamefont{Abeykoon}},
  \bibinfo{author}{\bibfnamefont{M.~J.} \bibnamefont{Krogstad}},
  \bibinfo{author}{\bibfnamefont{S.}~\bibnamefont{Rosenkranz}},
  \bibinfo{author}{\bibfnamefont{R.}~\bibnamefont{Osborn}},
  \bibnamefont{et~al.}, \bibinfo{journal}{Physical Review Letters}
  \textbf{\bibinfo{volume}{125}}, \bibinfo{pages}{247002}
  (\bibinfo{year}{2020}), ISSN \bibinfo{issn}{0031-9007}.

\bibitem[{\citenamefont{Jiang et~al.}(2021{\natexlab{a}})\citenamefont{Jiang,
  Yin, Denner, Shumiya, Ortiz, Xu, Guguchia, He, Hossain, Liu
  et~al.}}]{Jiang2021}
\bibinfo{author}{\bibfnamefont{Y.-X.} \bibnamefont{Jiang}},
  \bibinfo{author}{\bibfnamefont{J.-X.} \bibnamefont{Yin}},
  \bibinfo{author}{\bibfnamefont{M.~M.} \bibnamefont{Denner}},
  \bibinfo{author}{\bibfnamefont{N.}~\bibnamefont{Shumiya}},
  \bibinfo{author}{\bibfnamefont{B.~R.} \bibnamefont{Ortiz}},
  \bibinfo{author}{\bibfnamefont{G.}~\bibnamefont{Xu}},
  \bibinfo{author}{\bibfnamefont{Z.}~\bibnamefont{Guguchia}},
  \bibinfo{author}{\bibfnamefont{J.}~\bibnamefont{He}},
  \bibinfo{author}{\bibfnamefont{M.~S.} \bibnamefont{Hossain}},
  \bibinfo{author}{\bibfnamefont{X.}~\bibnamefont{Liu}}, \bibnamefont{et~al.},
  \bibinfo{journal}{Nature Materials} \textbf{\bibinfo{volume}{20}},
  \bibinfo{pages}{1353} (\bibinfo{year}{2021}{\natexlab{a}}), ISSN
  \bibinfo{issn}{1476-1122}.

\bibitem[{\citenamefont{Zhao et~al.}(2021)\citenamefont{Zhao, Li, Ortiz,
  Teicher, Park, Ye, Wang, Balents, Wilson, and Zeljkovic}}]{Zhao2021}
\bibinfo{author}{\bibfnamefont{H.}~\bibnamefont{Zhao}},
  \bibinfo{author}{\bibfnamefont{H.}~\bibnamefont{Li}},
  \bibinfo{author}{\bibfnamefont{B.~R.} \bibnamefont{Ortiz}},
  \bibinfo{author}{\bibfnamefont{S.~M.~L.} \bibnamefont{Teicher}},
  \bibinfo{author}{\bibfnamefont{T.}~\bibnamefont{Park}},
  \bibinfo{author}{\bibfnamefont{M.}~\bibnamefont{Ye}},
  \bibinfo{author}{\bibfnamefont{Z.}~\bibnamefont{Wang}},
  \bibinfo{author}{\bibfnamefont{L.}~\bibnamefont{Balents}},
  \bibinfo{author}{\bibfnamefont{S.~D.} \bibnamefont{Wilson}},
  \bibnamefont{and}
  \bibinfo{author}{\bibfnamefont{I.}~\bibnamefont{Zeljkovic}},
  \bibinfo{journal}{Nature} \textbf{\bibinfo{volume}{599}},
  \bibinfo{pages}{216} (\bibinfo{year}{2021}), ISSN \bibinfo{issn}{0028-0836}.

\bibitem[{\citenamefont{Ortiz et~al.}(2021{\natexlab{a}})\citenamefont{Ortiz,
  Sarte, Kenney, Graf, Teicher, Seshadri, and Wilson}}]{Ortiz2021}
\bibinfo{author}{\bibfnamefont{B.~R.} \bibnamefont{Ortiz}},
  \bibinfo{author}{\bibfnamefont{P.~M.} \bibnamefont{Sarte}},
  \bibinfo{author}{\bibfnamefont{E.~M.} \bibnamefont{Kenney}},
  \bibinfo{author}{\bibfnamefont{M.~J.} \bibnamefont{Graf}},
  \bibinfo{author}{\bibfnamefont{S.~M.~L.} \bibnamefont{Teicher}},
  \bibinfo{author}{\bibfnamefont{R.}~\bibnamefont{Seshadri}}, \bibnamefont{and}
  \bibinfo{author}{\bibfnamefont{S.~D.} \bibnamefont{Wilson}},
  \bibinfo{journal}{Physical Review Materials} \textbf{\bibinfo{volume}{5}},
  \bibinfo{pages}{034801} (\bibinfo{year}{2021}{\natexlab{a}}), ISSN
  \bibinfo{issn}{2475-9953}.

\bibitem[{\citenamefont{Chen et~al.}(2021{\natexlab{a}})\citenamefont{Chen,
  Wang, Yin, Gu, Jiang, Tu, Gong, Uwatoko, Sun, Lei et~al.}}]{Chen2021}
\bibinfo{author}{\bibfnamefont{K.}~\bibnamefont{Chen}},
  \bibinfo{author}{\bibfnamefont{N.}~\bibnamefont{Wang}},
  \bibinfo{author}{\bibfnamefont{Q.}~\bibnamefont{Yin}},
  \bibinfo{author}{\bibfnamefont{Y.}~\bibnamefont{Gu}},
  \bibinfo{author}{\bibfnamefont{K.}~\bibnamefont{Jiang}},
  \bibinfo{author}{\bibfnamefont{Z.}~\bibnamefont{Tu}},
  \bibinfo{author}{\bibfnamefont{C.}~\bibnamefont{Gong}},
  \bibinfo{author}{\bibfnamefont{Y.}~\bibnamefont{Uwatoko}},
  \bibinfo{author}{\bibfnamefont{J.}~\bibnamefont{Sun}},
  \bibinfo{author}{\bibfnamefont{H.}~\bibnamefont{Lei}}, \bibnamefont{et~al.},
  \bibinfo{journal}{Physical Review Letters} \textbf{\bibinfo{volume}{126}},
  \bibinfo{pages}{247001} (\bibinfo{year}{2021}{\natexlab{a}}), ISSN
  \bibinfo{issn}{0031-9007}.

\bibitem[{\citenamefont{Chen et~al.}(2021{\natexlab{b}})\citenamefont{Chen,
  Yang, Hu, Zhao, Yuan, Xing, Qian, Huang, Li, Ye et~al.}}]{ChenH2021}
\bibinfo{author}{\bibfnamefont{H.}~\bibnamefont{Chen}},
  \bibinfo{author}{\bibfnamefont{H.}~\bibnamefont{Yang}},
  \bibinfo{author}{\bibfnamefont{B.}~\bibnamefont{Hu}},
  \bibinfo{author}{\bibfnamefont{Z.}~\bibnamefont{Zhao}},
  \bibinfo{author}{\bibfnamefont{J.}~\bibnamefont{Yuan}},
  \bibinfo{author}{\bibfnamefont{Y.}~\bibnamefont{Xing}},
  \bibinfo{author}{\bibfnamefont{G.}~\bibnamefont{Qian}},
  \bibinfo{author}{\bibfnamefont{Z.}~\bibnamefont{Huang}},
  \bibinfo{author}{\bibfnamefont{G.}~\bibnamefont{Li}},
  \bibinfo{author}{\bibfnamefont{Y.}~\bibnamefont{Ye}}, \bibnamefont{et~al.},
  \bibinfo{journal}{Nature} \textbf{\bibinfo{volume}{599}},
  \bibinfo{pages}{222} (\bibinfo{year}{2021}{\natexlab{b}}), ISSN
  \bibinfo{issn}{0028-0836}.

\bibitem[{\citenamefont{Yu et~al.}(2021{\natexlab{a}})\citenamefont{Yu, Ma,
  Zhuo, Liu, Wen, Lei, Ying, and Chen}}]{Yu2021}
\bibinfo{author}{\bibfnamefont{F.~H.} \bibnamefont{Yu}},
  \bibinfo{author}{\bibfnamefont{D.~H.} \bibnamefont{Ma}},
  \bibinfo{author}{\bibfnamefont{W.~Z.} \bibnamefont{Zhuo}},
  \bibinfo{author}{\bibfnamefont{S.~Q.} \bibnamefont{Liu}},
  \bibinfo{author}{\bibfnamefont{X.~K.} \bibnamefont{Wen}},
  \bibinfo{author}{\bibfnamefont{B.}~\bibnamefont{Lei}},
  \bibinfo{author}{\bibfnamefont{J.~J.} \bibnamefont{Ying}}, \bibnamefont{and}
  \bibinfo{author}{\bibfnamefont{X.~H.} \bibnamefont{Chen}},
  \bibinfo{journal}{Nature Communications} \textbf{\bibinfo{volume}{12}},
  \bibinfo{pages}{3645} (\bibinfo{year}{2021}{\natexlab{a}}), ISSN
  \bibinfo{issn}{2041-1723}.

\bibitem[{\citenamefont{Xu et~al.}(2021)\citenamefont{Xu, Yan, Yin, Xia, Fang,
  Chen, Li, Yang, Guo, and Feng}}]{Xu2021}
\bibinfo{author}{\bibfnamefont{H.-S.} \bibnamefont{Xu}},
  \bibinfo{author}{\bibfnamefont{Y.-J.} \bibnamefont{Yan}},
  \bibinfo{author}{\bibfnamefont{R.}~\bibnamefont{Yin}},
  \bibinfo{author}{\bibfnamefont{W.}~\bibnamefont{Xia}},
  \bibinfo{author}{\bibfnamefont{S.}~\bibnamefont{Fang}},
  \bibinfo{author}{\bibfnamefont{Z.}~\bibnamefont{Chen}},
  \bibinfo{author}{\bibfnamefont{Y.}~\bibnamefont{Li}},
  \bibinfo{author}{\bibfnamefont{W.}~\bibnamefont{Yang}},
  \bibinfo{author}{\bibfnamefont{Y.}~\bibnamefont{Guo}}, \bibnamefont{and}
  \bibinfo{author}{\bibfnamefont{D.-L.} \bibnamefont{Feng}},
  \bibinfo{journal}{Physical Review Letters} \textbf{\bibinfo{volume}{127}},
  \bibinfo{pages}{187004} (\bibinfo{year}{2021}), ISSN
  \bibinfo{issn}{0031-9007}.

\bibitem[{\citenamefont{Yin et~al.}(2021)\citenamefont{Yin, Zhang, Chen, Ye,
  Yu, Ortiz, Luo, Duan, Su, Ying et~al.}}]{Yin2021}
\bibinfo{author}{\bibfnamefont{L.}~\bibnamefont{Yin}},
  \bibinfo{author}{\bibfnamefont{D.}~\bibnamefont{Zhang}},
  \bibinfo{author}{\bibfnamefont{C.}~\bibnamefont{Chen}},
  \bibinfo{author}{\bibfnamefont{G.}~\bibnamefont{Ye}},
  \bibinfo{author}{\bibfnamefont{F.}~\bibnamefont{Yu}},
  \bibinfo{author}{\bibfnamefont{B.~R.} \bibnamefont{Ortiz}},
  \bibinfo{author}{\bibfnamefont{S.}~\bibnamefont{Luo}},
  \bibinfo{author}{\bibfnamefont{W.}~\bibnamefont{Duan}},
  \bibinfo{author}{\bibfnamefont{H.}~\bibnamefont{Su}},
  \bibinfo{author}{\bibfnamefont{J.}~\bibnamefont{Ying}}, \bibnamefont{et~al.},
  \bibinfo{journal}{Physical Review B} \textbf{\bibinfo{volume}{104}},
  \bibinfo{pages}{174507} (\bibinfo{year}{2021}), ISSN
  \bibinfo{issn}{2469-9950}.

\bibitem[{\citenamefont{Li et~al.}(2021)\citenamefont{Li, Zhang, Yilmaz, Pai,
  Marvinney, Said, Yin, Gong, Tu, Vescovo et~al.}}]{Li2021}
\bibinfo{author}{\bibfnamefont{H.}~\bibnamefont{Li}},
  \bibinfo{author}{\bibfnamefont{T.}~\bibnamefont{Zhang}},
  \bibinfo{author}{\bibfnamefont{T.}~\bibnamefont{Yilmaz}},
  \bibinfo{author}{\bibfnamefont{Y.}~\bibnamefont{Pai}},
  \bibinfo{author}{\bibfnamefont{C.}~\bibnamefont{Marvinney}},
  \bibinfo{author}{\bibfnamefont{A.}~\bibnamefont{Said}},
  \bibinfo{author}{\bibfnamefont{Q.}~\bibnamefont{Yin}},
  \bibinfo{author}{\bibfnamefont{C.}~\bibnamefont{Gong}},
  \bibinfo{author}{\bibfnamefont{Z.}~\bibnamefont{Tu}},
  \bibinfo{author}{\bibfnamefont{E.}~\bibnamefont{Vescovo}},
  \bibnamefont{et~al.}, \bibinfo{journal}{Physical Review X}
  \textbf{\bibinfo{volume}{11}}, \bibinfo{pages}{031050}
  (\bibinfo{year}{2021}), ISSN \bibinfo{issn}{2160-3308}.

\bibitem[{\citenamefont{Zhang et~al.}(2021)\citenamefont{Zhang, Chen, Zhou,
  Yuan, Wang, Wang, Yang, An, Zhang, Zhu et~al.}}]{Zhang2021}
\bibinfo{author}{\bibfnamefont{Z.}~\bibnamefont{Zhang}},
  \bibinfo{author}{\bibfnamefont{Z.}~\bibnamefont{Chen}},
  \bibinfo{author}{\bibfnamefont{Y.}~\bibnamefont{Zhou}},
  \bibinfo{author}{\bibfnamefont{Y.}~\bibnamefont{Yuan}},
  \bibinfo{author}{\bibfnamefont{S.}~\bibnamefont{Wang}},
  \bibinfo{author}{\bibfnamefont{J.}~\bibnamefont{Wang}},
  \bibinfo{author}{\bibfnamefont{H.}~\bibnamefont{Yang}},
  \bibinfo{author}{\bibfnamefont{C.}~\bibnamefont{An}},
  \bibinfo{author}{\bibfnamefont{L.}~\bibnamefont{Zhang}},
  \bibinfo{author}{\bibfnamefont{X.}~\bibnamefont{Zhu}}, \bibnamefont{et~al.},
  \bibinfo{journal}{Physical Review B} \textbf{\bibinfo{volume}{103}},
  \bibinfo{pages}{224513} (\bibinfo{year}{2021}), ISSN
  \bibinfo{issn}{2469-9950}.

\bibitem[{\citenamefont{Ortiz et~al.}(2021{\natexlab{b}})\citenamefont{Ortiz,
  Teicher, Kautzsch, Sarte, Ratcliff, Harter, Ruff, Seshadri, and
  Wilson}}]{Ortiz2021PRX}
\bibinfo{author}{\bibfnamefont{B.~R.} \bibnamefont{Ortiz}},
  \bibinfo{author}{\bibfnamefont{S.~M.} \bibnamefont{Teicher}},
  \bibinfo{author}{\bibfnamefont{L.}~\bibnamefont{Kautzsch}},
  \bibinfo{author}{\bibfnamefont{P.~M.} \bibnamefont{Sarte}},
  \bibinfo{author}{\bibfnamefont{N.}~\bibnamefont{Ratcliff}},
  \bibinfo{author}{\bibfnamefont{J.}~\bibnamefont{Harter}},
  \bibinfo{author}{\bibfnamefont{J.~P.} \bibnamefont{Ruff}},
  \bibinfo{author}{\bibfnamefont{R.}~\bibnamefont{Seshadri}}, \bibnamefont{and}
  \bibinfo{author}{\bibfnamefont{S.~D.} \bibnamefont{Wilson}},
  \bibinfo{journal}{Physical Review X} \textbf{\bibinfo{volume}{11}},
  \bibinfo{pages}{041030} (\bibinfo{year}{2021}{\natexlab{b}}), ISSN
  \bibinfo{issn}{2160-3308}.

\bibitem[{\citenamefont{Wu et~al.}(2021{\natexlab{b}})\citenamefont{Wu, Wang,
  Liu, Li, Xu, Yin, Gong, Tu, Lei, Dong et~al.}}]{Wu2021}
\bibinfo{author}{\bibfnamefont{Q.}~\bibnamefont{Wu}},
  \bibinfo{author}{\bibfnamefont{Z.~X.} \bibnamefont{Wang}},
  \bibinfo{author}{\bibfnamefont{Q.~M.} \bibnamefont{Liu}},
  \bibinfo{author}{\bibfnamefont{R.~S.} \bibnamefont{Li}},
  \bibinfo{author}{\bibfnamefont{S.~X.} \bibnamefont{Xu}},
  \bibinfo{author}{\bibfnamefont{Q.~W.} \bibnamefont{Yin}},
  \bibinfo{author}{\bibfnamefont{C.~S.} \bibnamefont{Gong}},
  \bibinfo{author}{\bibfnamefont{Z.~J.} \bibnamefont{Tu}},
  \bibinfo{author}{\bibfnamefont{H.~C.} \bibnamefont{Lei}},
  \bibinfo{author}{\bibfnamefont{T.}~\bibnamefont{Dong}}, \bibnamefont{et~al.}
  (\bibinfo{year}{2021}{\natexlab{b}}).

\bibitem[{\citenamefont{Kang et~al.}(2021)\citenamefont{Kang, Fang, Kim, Ortiz,
  Ryu, Kim, Yoo, Sangiovanni, Sante, Park et~al.}}]{Kang2021}
\bibinfo{author}{\bibfnamefont{M.}~\bibnamefont{Kang}},
  \bibinfo{author}{\bibfnamefont{S.}~\bibnamefont{Fang}},
  \bibinfo{author}{\bibfnamefont{J.-K.} \bibnamefont{Kim}},
  \bibinfo{author}{\bibfnamefont{B.~R.} \bibnamefont{Ortiz}},
  \bibinfo{author}{\bibfnamefont{S.~H.} \bibnamefont{Ryu}},
  \bibinfo{author}{\bibfnamefont{J.}~\bibnamefont{Kim}},
  \bibinfo{author}{\bibfnamefont{J.}~\bibnamefont{Yoo}},
  \bibinfo{author}{\bibfnamefont{G.}~\bibnamefont{Sangiovanni}},
  \bibinfo{author}{\bibfnamefont{D.~D.} \bibnamefont{Sante}},
  \bibinfo{author}{\bibfnamefont{B.-G.} \bibnamefont{Park}},
  \bibnamefont{et~al.} (\bibinfo{year}{2021}).

\bibitem[{\citenamefont{Yu et~al.}(2021{\natexlab{b}})\citenamefont{Yu, Wang,
  Zhang, Sander, Ni, Lu, Ma, Wang, Zhao, Chen et~al.}}]{yu2021evidence}
\bibinfo{author}{\bibfnamefont{L.}~\bibnamefont{Yu}},
  \bibinfo{author}{\bibfnamefont{C.}~\bibnamefont{Wang}},
  \bibinfo{author}{\bibfnamefont{Y.}~\bibnamefont{Zhang}},
  \bibinfo{author}{\bibfnamefont{M.}~\bibnamefont{Sander}},
  \bibinfo{author}{\bibfnamefont{S.}~\bibnamefont{Ni}},
  \bibinfo{author}{\bibfnamefont{Z.}~\bibnamefont{Lu}},
  \bibinfo{author}{\bibfnamefont{S.}~\bibnamefont{Ma}},
  \bibinfo{author}{\bibfnamefont{Z.}~\bibnamefont{Wang}},
  \bibinfo{author}{\bibfnamefont{Z.}~\bibnamefont{Zhao}},
  \bibinfo{author}{\bibfnamefont{H.}~\bibnamefont{Chen}}, \bibnamefont{et~al.},
  \bibinfo{journal}{arXiv preprint arXiv:2107.10714}
  (\bibinfo{year}{2021}{\natexlab{b}}).

\bibitem[{\citenamefont{Nakayama et~al.}(2021)\citenamefont{Nakayama, Li, Kato,
  Liu, Wang, Takahashi, Yao, and Sato}}]{Nakayama2021}
\bibinfo{author}{\bibfnamefont{K.}~\bibnamefont{Nakayama}},
  \bibinfo{author}{\bibfnamefont{Y.}~\bibnamefont{Li}},
  \bibinfo{author}{\bibfnamefont{T.}~\bibnamefont{Kato}},
  \bibinfo{author}{\bibfnamefont{M.}~\bibnamefont{Liu}},
  \bibinfo{author}{\bibfnamefont{Z.}~\bibnamefont{Wang}},
  \bibinfo{author}{\bibfnamefont{T.}~\bibnamefont{Takahashi}},
  \bibinfo{author}{\bibfnamefont{Y.}~\bibnamefont{Yao}}, \bibnamefont{and}
  \bibinfo{author}{\bibfnamefont{T.}~\bibnamefont{Sato}},
  \bibinfo{journal}{Physical Review B} \textbf{\bibinfo{volume}{104}},
  \bibinfo{pages}{L161112} (\bibinfo{year}{2021}), ISSN
  \bibinfo{issn}{2469-9950}.

\bibitem[{\citenamefont{Li et~al.}(2022{\natexlab{a}})\citenamefont{Li, Wan,
  Li, Li, Gu, Yang, Li, Wang, Yao, and Wen}}]{Li2022}
\bibinfo{author}{\bibfnamefont{H.}~\bibnamefont{Li}},
  \bibinfo{author}{\bibfnamefont{S.}~\bibnamefont{Wan}},
  \bibinfo{author}{\bibfnamefont{H.}~\bibnamefont{Li}},
  \bibinfo{author}{\bibfnamefont{Q.}~\bibnamefont{Li}},
  \bibinfo{author}{\bibfnamefont{Q.}~\bibnamefont{Gu}},
  \bibinfo{author}{\bibfnamefont{H.}~\bibnamefont{Yang}},
  \bibinfo{author}{\bibfnamefont{Y.}~\bibnamefont{Li}},
  \bibinfo{author}{\bibfnamefont{Z.}~\bibnamefont{Wang}},
  \bibinfo{author}{\bibfnamefont{Y.}~\bibnamefont{Yao}}, \bibnamefont{and}
  \bibinfo{author}{\bibfnamefont{H.-H.} \bibnamefont{Wen}},
  \bibinfo{journal}{Physical Review B} \textbf{\bibinfo{volume}{105}},
  \bibinfo{pages}{045102} (\bibinfo{year}{2022}{\natexlab{a}}), ISSN
  \bibinfo{issn}{2469-9950}.

\bibitem[{\citenamefont{Liu et~al.}(2021)\citenamefont{Liu, Zhao, Yin, Gong,
  Tu, Li, Song, Liu, Shen, Huang et~al.}}]{Liu2021}
\bibinfo{author}{\bibfnamefont{Z.}~\bibnamefont{Liu}},
  \bibinfo{author}{\bibfnamefont{N.}~\bibnamefont{Zhao}},
  \bibinfo{author}{\bibfnamefont{Q.}~\bibnamefont{Yin}},
  \bibinfo{author}{\bibfnamefont{C.}~\bibnamefont{Gong}},
  \bibinfo{author}{\bibfnamefont{Z.}~\bibnamefont{Tu}},
  \bibinfo{author}{\bibfnamefont{M.}~\bibnamefont{Li}},
  \bibinfo{author}{\bibfnamefont{W.}~\bibnamefont{Song}},
  \bibinfo{author}{\bibfnamefont{Z.}~\bibnamefont{Liu}},
  \bibinfo{author}{\bibfnamefont{D.}~\bibnamefont{Shen}},
  \bibinfo{author}{\bibfnamefont{Y.}~\bibnamefont{Huang}},
  \bibnamefont{et~al.}, \bibinfo{journal}{Physical Review X}
  \textbf{\bibinfo{volume}{11}}, \bibinfo{pages}{041010}
  (\bibinfo{year}{2021}), ISSN \bibinfo{issn}{2160-3308}.

\bibitem[{\citenamefont{Cho et~al.}(2021)\citenamefont{Cho, Ma, Xia, Yang, Liu,
  Huang, Jiang, Lu, Liu, Liu et~al.}}]{Cho2021}
\bibinfo{author}{\bibfnamefont{S.}~\bibnamefont{Cho}},
  \bibinfo{author}{\bibfnamefont{H.}~\bibnamefont{Ma}},
  \bibinfo{author}{\bibfnamefont{W.}~\bibnamefont{Xia}},
  \bibinfo{author}{\bibfnamefont{Y.}~\bibnamefont{Yang}},
  \bibinfo{author}{\bibfnamefont{Z.}~\bibnamefont{Liu}},
  \bibinfo{author}{\bibfnamefont{Z.}~\bibnamefont{Huang}},
  \bibinfo{author}{\bibfnamefont{Z.}~\bibnamefont{Jiang}},
  \bibinfo{author}{\bibfnamefont{X.}~\bibnamefont{Lu}},
  \bibinfo{author}{\bibfnamefont{J.}~\bibnamefont{Liu}},
  \bibinfo{author}{\bibfnamefont{Z.}~\bibnamefont{Liu}}, \bibnamefont{et~al.},
  \bibinfo{journal}{Physical Review Letters} \textbf{\bibinfo{volume}{127}},
  \bibinfo{pages}{236401} (\bibinfo{year}{2021}), ISSN
  \bibinfo{issn}{0031-9007}.

\bibitem[{\citenamefont{Zhou et~al.}(2021)\citenamefont{Zhou, Li, Fan, Hao,
  Dai, Wang, Yao, and Wen}}]{Zhou2021}
\bibinfo{author}{\bibfnamefont{X.}~\bibnamefont{Zhou}},
  \bibinfo{author}{\bibfnamefont{Y.}~\bibnamefont{Li}},
  \bibinfo{author}{\bibfnamefont{X.}~\bibnamefont{Fan}},
  \bibinfo{author}{\bibfnamefont{J.}~\bibnamefont{Hao}},
  \bibinfo{author}{\bibfnamefont{Y.}~\bibnamefont{Dai}},
  \bibinfo{author}{\bibfnamefont{Z.}~\bibnamefont{Wang}},
  \bibinfo{author}{\bibfnamefont{Y.}~\bibnamefont{Yao}}, \bibnamefont{and}
  \bibinfo{author}{\bibfnamefont{H.-H.} \bibnamefont{Wen}},
  \bibinfo{journal}{Physical Review B} \textbf{\bibinfo{volume}{104}},
  \bibinfo{pages}{L041101} (\bibinfo{year}{2021}), ISSN
  \bibinfo{issn}{2469-9950}.

\bibitem[{\citenamefont{Xie et~al.}(2022)\citenamefont{Xie, Li, Bourges,
  Ivanov, Ye, Yin, Hasan, Luo, Yao, Wang et~al.}}]{Xie2022}
\bibinfo{author}{\bibfnamefont{Y.}~\bibnamefont{Xie}},
  \bibinfo{author}{\bibfnamefont{Y.}~\bibnamefont{Li}},
  \bibinfo{author}{\bibfnamefont{P.}~\bibnamefont{Bourges}},
  \bibinfo{author}{\bibfnamefont{A.}~\bibnamefont{Ivanov}},
  \bibinfo{author}{\bibfnamefont{Z.}~\bibnamefont{Ye}},
  \bibinfo{author}{\bibfnamefont{J.-X.} \bibnamefont{Yin}},
  \bibinfo{author}{\bibfnamefont{M.~Z.} \bibnamefont{Hasan}},
  \bibinfo{author}{\bibfnamefont{A.}~\bibnamefont{Luo}},
  \bibinfo{author}{\bibfnamefont{Y.}~\bibnamefont{Yao}},
  \bibinfo{author}{\bibfnamefont{Z.}~\bibnamefont{Wang}}, \bibnamefont{et~al.},
  \bibinfo{journal}{Physical Review B} \textbf{\bibinfo{volume}{105}},
  \bibinfo{pages}{L140501} (\bibinfo{year}{2022}), ISSN
  \bibinfo{issn}{2469-9950}.

\bibitem[{\citenamefont{Wulferding et~al.}(2021)\citenamefont{Wulferding, Lee,
  Choi, Yin, Tu, Gong, Lei, and Choi}}]{Wulferding2021}
\bibinfo{author}{\bibfnamefont{D.}~\bibnamefont{Wulferding}},
  \bibinfo{author}{\bibfnamefont{S.}~\bibnamefont{Lee}},
  \bibinfo{author}{\bibfnamefont{Y.}~\bibnamefont{Choi}},
  \bibinfo{author}{\bibfnamefont{Q.}~\bibnamefont{Yin}},
  \bibinfo{author}{\bibfnamefont{Z.}~\bibnamefont{Tu}},
  \bibinfo{author}{\bibfnamefont{C.}~\bibnamefont{Gong}},
  \bibinfo{author}{\bibfnamefont{H.}~\bibnamefont{Lei}}, \bibnamefont{and}
  \bibinfo{author}{\bibfnamefont{K.-Y.} \bibnamefont{Choi}}
  (\bibinfo{year}{2021}).

\bibitem[{\citenamefont{Wang et~al.}(2021)\citenamefont{Wang, Wu, Yin, Gong,
  Tu, Lin, Liu, Shi, Zhang, Wu et~al.}}]{Wang2021}
\bibinfo{author}{\bibfnamefont{Z.~X.} \bibnamefont{Wang}},
  \bibinfo{author}{\bibfnamefont{Q.}~\bibnamefont{Wu}},
  \bibinfo{author}{\bibfnamefont{Q.~W.} \bibnamefont{Yin}},
  \bibinfo{author}{\bibfnamefont{C.~S.} \bibnamefont{Gong}},
  \bibinfo{author}{\bibfnamefont{Z.~J.} \bibnamefont{Tu}},
  \bibinfo{author}{\bibfnamefont{T.}~\bibnamefont{Lin}},
  \bibinfo{author}{\bibfnamefont{Q.~M.} \bibnamefont{Liu}},
  \bibinfo{author}{\bibfnamefont{L.~Y.} \bibnamefont{Shi}},
  \bibinfo{author}{\bibfnamefont{S.~J.} \bibnamefont{Zhang}},
  \bibinfo{author}{\bibfnamefont{D.}~\bibnamefont{Wu}}, \bibnamefont{et~al.},
  \bibinfo{journal}{Physical Review B} \textbf{\bibinfo{volume}{104}},
  \bibinfo{pages}{165110} (\bibinfo{year}{2021}), ISSN
  \bibinfo{issn}{2469-9950}.

\bibitem[{\citenamefont{Song et~al.}(2021)\citenamefont{Song, Kong, Xia, Yin,
  Tu, Zhao, Dai, Meng, Tao, Tu et~al.}}]{Song2021}
\bibinfo{author}{\bibfnamefont{B.~Q.} \bibnamefont{Song}},
  \bibinfo{author}{\bibfnamefont{X.~M.} \bibnamefont{Kong}},
  \bibinfo{author}{\bibfnamefont{W.}~\bibnamefont{Xia}},
  \bibinfo{author}{\bibfnamefont{Q.~W.} \bibnamefont{Yin}},
  \bibinfo{author}{\bibfnamefont{C.~P.} \bibnamefont{Tu}},
  \bibinfo{author}{\bibfnamefont{C.~C.} \bibnamefont{Zhao}},
  \bibinfo{author}{\bibfnamefont{D.~Z.} \bibnamefont{Dai}},
  \bibinfo{author}{\bibfnamefont{K.}~\bibnamefont{Meng}},
  \bibinfo{author}{\bibfnamefont{Z.~C.} \bibnamefont{Tao}},
  \bibinfo{author}{\bibfnamefont{Z.~J.} \bibnamefont{Tu}}, \bibnamefont{et~al.}
  (\bibinfo{year}{2021}).

\bibitem[{\citenamefont{Nie et~al.}(2022)\citenamefont{Nie, Sun, Ma, Song,
  Zheng, Liang, Wu, Yu, Li, Shan et~al.}}]{Nie2022}
\bibinfo{author}{\bibfnamefont{L.}~\bibnamefont{Nie}},
  \bibinfo{author}{\bibfnamefont{K.}~\bibnamefont{Sun}},
  \bibinfo{author}{\bibfnamefont{W.}~\bibnamefont{Ma}},
  \bibinfo{author}{\bibfnamefont{D.}~\bibnamefont{Song}},
  \bibinfo{author}{\bibfnamefont{L.}~\bibnamefont{Zheng}},
  \bibinfo{author}{\bibfnamefont{Z.}~\bibnamefont{Liang}},
  \bibinfo{author}{\bibfnamefont{P.}~\bibnamefont{Wu}},
  \bibinfo{author}{\bibfnamefont{F.}~\bibnamefont{Yu}},
  \bibinfo{author}{\bibfnamefont{J.}~\bibnamefont{Li}},
  \bibinfo{author}{\bibfnamefont{M.}~\bibnamefont{Shan}}, \bibnamefont{et~al.},
  \bibinfo{journal}{Nature} \textbf{\bibinfo{volume}{604}}, \bibinfo{pages}{59}
  (\bibinfo{year}{2022}), ISSN \bibinfo{issn}{0028-0836}.

\bibitem[{\citenamefont{Mielke et~al.}(2022)\citenamefont{Mielke, Das, Yin,
  Liu, Gupta, Jiang, Medarde, Wu, Lei, Chang et~al.}}]{mielke-muon-nature22}
\bibinfo{author}{\bibfnamefont{C.}~\bibnamefont{Mielke}},
  \bibinfo{author}{\bibfnamefont{D.}~\bibnamefont{Das}},
  \bibinfo{author}{\bibfnamefont{J.-X.} \bibnamefont{Yin}},
  \bibinfo{author}{\bibfnamefont{H.}~\bibnamefont{Liu}},
  \bibinfo{author}{\bibfnamefont{R.}~\bibnamefont{Gupta}},
  \bibinfo{author}{\bibfnamefont{Y.-X.} \bibnamefont{Jiang}},
  \bibinfo{author}{\bibfnamefont{M.}~\bibnamefont{Medarde}},
  \bibinfo{author}{\bibfnamefont{X.}~\bibnamefont{Wu}},
  \bibinfo{author}{\bibfnamefont{H.~C.} \bibnamefont{Lei}},
  \bibinfo{author}{\bibfnamefont{J.}~\bibnamefont{Chang}},
  \bibnamefont{et~al.}, \bibinfo{journal}{Nature}
  \textbf{\bibinfo{volume}{602}}, \bibinfo{pages}{245} (\bibinfo{year}{2022}),
  ISSN \bibinfo{issn}{1476-4687}.

\bibitem[{\citenamefont{Lou et~al.}(2022)\citenamefont{Lou, Fedorov, Yin,
  Kuibarov, Tu, Gong, Schwier, Büchner, Lei, and Borisenko}}]{Lou2022}
\bibinfo{author}{\bibfnamefont{R.}~\bibnamefont{Lou}},
  \bibinfo{author}{\bibfnamefont{A.}~\bibnamefont{Fedorov}},
  \bibinfo{author}{\bibfnamefont{Q.}~\bibnamefont{Yin}},
  \bibinfo{author}{\bibfnamefont{A.}~\bibnamefont{Kuibarov}},
  \bibinfo{author}{\bibfnamefont{Z.}~\bibnamefont{Tu}},
  \bibinfo{author}{\bibfnamefont{C.}~\bibnamefont{Gong}},
  \bibinfo{author}{\bibfnamefont{E.~F.} \bibnamefont{Schwier}},
  \bibinfo{author}{\bibfnamefont{B.}~\bibnamefont{Büchner}},
  \bibinfo{author}{\bibfnamefont{H.}~\bibnamefont{Lei}}, \bibnamefont{and}
  \bibinfo{author}{\bibfnamefont{S.}~\bibnamefont{Borisenko}},
  \bibinfo{journal}{Physical Review Letters} \textbf{\bibinfo{volume}{128}},
  \bibinfo{pages}{036402} (\bibinfo{year}{2022}), ISSN
  \bibinfo{issn}{0031-9007}.

\bibitem[{\citenamefont{Li et~al.}(2022{\natexlab{b}})\citenamefont{Li, Zhao,
  Ortiz, Park, Ye, Balents, Wang, Wilson, and Zeljkovic}}]{Li2022np}
\bibinfo{author}{\bibfnamefont{H.}~\bibnamefont{Li}},
  \bibinfo{author}{\bibfnamefont{H.}~\bibnamefont{Zhao}},
  \bibinfo{author}{\bibfnamefont{B.~R.} \bibnamefont{Ortiz}},
  \bibinfo{author}{\bibfnamefont{T.}~\bibnamefont{Park}},
  \bibinfo{author}{\bibfnamefont{M.}~\bibnamefont{Ye}},
  \bibinfo{author}{\bibfnamefont{L.}~\bibnamefont{Balents}},
  \bibinfo{author}{\bibfnamefont{Z.}~\bibnamefont{Wang}},
  \bibinfo{author}{\bibfnamefont{S.~D.} \bibnamefont{Wilson}},
  \bibnamefont{and}
  \bibinfo{author}{\bibfnamefont{I.}~\bibnamefont{Zeljkovic}},
  \bibinfo{journal}{Nature Physics} \textbf{\bibinfo{volume}{18}},
  \bibinfo{pages}{265} (\bibinfo{year}{2022}{\natexlab{b}}), ISSN
  \bibinfo{issn}{1745-2473}.

\bibitem[{\citenamefont{Wu et~al.}(2022)\citenamefont{Wu, Ortiz, Tan, Wilson,
  Yan, Birol, and Blumberg}}]{Wu2022}
\bibinfo{author}{\bibfnamefont{S.}~\bibnamefont{Wu}},
  \bibinfo{author}{\bibfnamefont{B.~R.} \bibnamefont{Ortiz}},
  \bibinfo{author}{\bibfnamefont{H.}~\bibnamefont{Tan}},
  \bibinfo{author}{\bibfnamefont{S.~D.} \bibnamefont{Wilson}},
  \bibinfo{author}{\bibfnamefont{B.}~\bibnamefont{Yan}},
  \bibinfo{author}{\bibfnamefont{T.}~\bibnamefont{Birol}}, \bibnamefont{and}
  \bibinfo{author}{\bibfnamefont{G.}~\bibnamefont{Blumberg}},
  \bibinfo{journal}{Physical Review B} \textbf{\bibinfo{volume}{105}},
  \bibinfo{pages}{155106} (\bibinfo{year}{2022}), ISSN
  \bibinfo{issn}{2469-9950}.

\bibitem[{\citenamefont{Shan et~al.}(2022)\citenamefont{Shan, Biswas, Ghosh,
  Tula, Hillier, Adroja, Cottrell, Cao, Liu, Xu et~al.}}]{shan-muon-prr22}
\bibinfo{author}{\bibfnamefont{Z.}~\bibnamefont{Shan}},
  \bibinfo{author}{\bibfnamefont{P.~K.} \bibnamefont{Biswas}},
  \bibinfo{author}{\bibfnamefont{S.~K.} \bibnamefont{Ghosh}},
  \bibinfo{author}{\bibfnamefont{T.}~\bibnamefont{Tula}},
  \bibinfo{author}{\bibfnamefont{A.~D.} \bibnamefont{Hillier}},
  \bibinfo{author}{\bibfnamefont{D.}~\bibnamefont{Adroja}},
  \bibinfo{author}{\bibfnamefont{S.}~\bibnamefont{Cottrell}},
  \bibinfo{author}{\bibfnamefont{G.-H.} \bibnamefont{Cao}},
  \bibinfo{author}{\bibfnamefont{Y.}~\bibnamefont{Liu}},
  \bibinfo{author}{\bibfnamefont{X.}~\bibnamefont{Xu}}, \bibnamefont{et~al.},
  \bibinfo{journal}{Phys. Rev. Research} \textbf{\bibinfo{volume}{4}},
  \bibinfo{pages}{033145} (\bibinfo{year}{2022}).

\bibitem[{\citenamefont{Jiang et~al.}(2021{\natexlab{b}})\citenamefont{Jiang,
  Yin, Denner, Shumiya, Ortiz, Xu, Guguchia, He, Hossain, Liu
  et~al.}}]{jiang2021unconventional}
\bibinfo{author}{\bibfnamefont{Y.-X.} \bibnamefont{Jiang}},
  \bibinfo{author}{\bibfnamefont{J.-X.} \bibnamefont{Yin}},
  \bibinfo{author}{\bibfnamefont{M.~M.} \bibnamefont{Denner}},
  \bibinfo{author}{\bibfnamefont{N.}~\bibnamefont{Shumiya}},
  \bibinfo{author}{\bibfnamefont{B.~R.} \bibnamefont{Ortiz}},
  \bibinfo{author}{\bibfnamefont{G.}~\bibnamefont{Xu}},
  \bibinfo{author}{\bibfnamefont{Z.}~\bibnamefont{Guguchia}},
  \bibinfo{author}{\bibfnamefont{J.}~\bibnamefont{He}},
  \bibinfo{author}{\bibfnamefont{M.~S.} \bibnamefont{Hossain}},
  \bibinfo{author}{\bibfnamefont{X.}~\bibnamefont{Liu}}, \bibnamefont{et~al.},
  \bibinfo{journal}{Nature Materials} pp. \bibinfo{pages}{1--5}
  (\bibinfo{year}{2021}{\natexlab{b}}).

\bibitem[{\citenamefont{Varma}(1997)}]{varma-prb97}
\bibinfo{author}{\bibfnamefont{C.~M.} \bibnamefont{Varma}},
  \bibinfo{journal}{Phys. Rev. B} \textbf{\bibinfo{volume}{55}},
  \bibinfo{pages}{14554} (\bibinfo{year}{1997}).

\bibitem[{\citenamefont{Liu et~al.}(2016)\citenamefont{Liu, Park, Garrity, and
  Vanderbilt}}]{liu-prl16}
\bibinfo{author}{\bibfnamefont{J.}~\bibnamefont{Liu}},
  \bibinfo{author}{\bibfnamefont{S.~Y.} \bibnamefont{Park}},
  \bibinfo{author}{\bibfnamefont{K.~F.} \bibnamefont{Garrity}},
  \bibnamefont{and}
  \bibinfo{author}{\bibfnamefont{D.}~\bibnamefont{Vanderbilt}},
  \bibinfo{journal}{Phys. Rev. Lett.} \textbf{\bibinfo{volume}{117}},
  \bibinfo{pages}{257201} (\bibinfo{year}{2016}),
  \urlprefix\url{https://link.aps.org/doi/10.1103/PhysRevLett.117.257201}.

\bibitem[{\citenamefont{Liu and Dai}(2021)}]{liu-nrp21}
\bibinfo{author}{\bibfnamefont{J.}~\bibnamefont{Liu}} \bibnamefont{and}
  \bibinfo{author}{\bibfnamefont{X.}~\bibnamefont{Dai}},
  \bibinfo{journal}{Nature Reviews Physics} \textbf{\bibinfo{volume}{3}},
  \bibinfo{pages}{367} (\bibinfo{year}{2021}).

\bibitem[{\citenamefont{Bourges et~al.}(2021)\citenamefont{Bourges, Bounoua,
  and Sidis}}]{bourges-review21}
\bibinfo{author}{\bibfnamefont{P.}~\bibnamefont{Bourges}},
  \bibinfo{author}{\bibfnamefont{D.}~\bibnamefont{Bounoua}}, \bibnamefont{and}
  \bibinfo{author}{\bibfnamefont{Y.}~\bibnamefont{Sidis}},
  \bibinfo{journal}{Comptes Rendus. Physique} \textbf{\bibinfo{volume}{22}},
  \bibinfo{pages}{7} (\bibinfo{year}{2021}).

\bibitem[{\citenamefont{Denner et~al.}(2021)\citenamefont{Denner, Thomale, and
  Neupert}}]{denner-prl21}
\bibinfo{author}{\bibfnamefont{M.~M.} \bibnamefont{Denner}},
  \bibinfo{author}{\bibfnamefont{R.}~\bibnamefont{Thomale}}, \bibnamefont{and}
  \bibinfo{author}{\bibfnamefont{T.}~\bibnamefont{Neupert}},
  \bibinfo{journal}{Phys. Rev. Lett.} \textbf{\bibinfo{volume}{127}},
  \bibinfo{pages}{217601} (\bibinfo{year}{2021}).

\bibitem[{\citenamefont{Ma and Liu}(2021)}]{Ma2021}
\bibinfo{author}{\bibfnamefont{H.-Y.} \bibnamefont{Ma}} \bibnamefont{and}
  \bibinfo{author}{\bibfnamefont{J.}~\bibnamefont{Liu}},
  \bibinfo{journal}{arXiv preprint arXiv:2112.02808}  (\bibinfo{year}{2021}).

\bibitem[{\citenamefont{{Yin} et~al.}(2022)\citenamefont{{Yin}, {Jiang},
  {Teng}, {Shafayat Hossain}, {Mardanya}, {Chang}, {Ye}, {Xu}, {Denner},
  {Neupert} et~al.}}]{yin-arxiv22}
\bibinfo{author}{\bibfnamefont{J.-X.} \bibnamefont{{Yin}}},
  \bibinfo{author}{\bibfnamefont{Y.-X.} \bibnamefont{{Jiang}}},
  \bibinfo{author}{\bibfnamefont{X.}~\bibnamefont{{Teng}}},
  \bibinfo{author}{\bibfnamefont{M.}~\bibnamefont{{Shafayat Hossain}}},
  \bibinfo{author}{\bibfnamefont{S.}~\bibnamefont{{Mardanya}}},
  \bibinfo{author}{\bibfnamefont{T.-R.} \bibnamefont{{Chang}}},
  \bibinfo{author}{\bibfnamefont{Z.}~\bibnamefont{{Ye}}},
  \bibinfo{author}{\bibfnamefont{G.}~\bibnamefont{{Xu}}},
  \bibinfo{author}{\bibfnamefont{M.~M.} \bibnamefont{{Denner}}},
  \bibinfo{author}{\bibfnamefont{T.}~\bibnamefont{{Neupert}}},
  \bibnamefont{et~al.}, \bibinfo{journal}{arXiv e-prints}
  \bibinfo{eid}{arXiv:2203.01888} (\bibinfo{year}{2022}), \eprint{2203.01888}.

\bibitem[{\citenamefont{Teng et~al.}(2022{\natexlab{a}})\citenamefont{Teng,
  Chen, Ye, Rosenberg, Liu, Yin, Jiang, Oh, Hasan, Neubauer
  et~al.}}]{teng2022discovery}
\bibinfo{author}{\bibfnamefont{X.}~\bibnamefont{Teng}},
  \bibinfo{author}{\bibfnamefont{L.}~\bibnamefont{Chen}},
  \bibinfo{author}{\bibfnamefont{F.}~\bibnamefont{Ye}},
  \bibinfo{author}{\bibfnamefont{E.}~\bibnamefont{Rosenberg}},
  \bibinfo{author}{\bibfnamefont{Z.}~\bibnamefont{Liu}},
  \bibinfo{author}{\bibfnamefont{J.-X.} \bibnamefont{Yin}},
  \bibinfo{author}{\bibfnamefont{Y.-X.} \bibnamefont{Jiang}},
  \bibinfo{author}{\bibfnamefont{J.~S.} \bibnamefont{Oh}},
  \bibinfo{author}{\bibfnamefont{M.~Z.} \bibnamefont{Hasan}},
  \bibinfo{author}{\bibfnamefont{K.~J.} \bibnamefont{Neubauer}},
  \bibnamefont{et~al.}, \bibinfo{journal}{Nature}
  (\bibinfo{year}{2022}{\natexlab{a}}).

\bibitem[{\citenamefont{Teng et~al.}(2022{\natexlab{b}})\citenamefont{Teng, Oh,
  Tan, Chen, Huang, Gao, Yin, Chu, Hashimoto, Lu et~al.}}]{fege-arpes-arxiv22}
\bibinfo{author}{\bibfnamefont{X.}~\bibnamefont{Teng}},
  \bibinfo{author}{\bibfnamefont{J.~S.} \bibnamefont{Oh}},
  \bibinfo{author}{\bibfnamefont{H.}~\bibnamefont{Tan}},
  \bibinfo{author}{\bibfnamefont{L.}~\bibnamefont{Chen}},
  \bibinfo{author}{\bibfnamefont{J.}~\bibnamefont{Huang}},
  \bibinfo{author}{\bibfnamefont{B.}~\bibnamefont{Gao}},
  \bibinfo{author}{\bibfnamefont{J.-X.} \bibnamefont{Yin}},
  \bibinfo{author}{\bibfnamefont{J.-H.} \bibnamefont{Chu}},
  \bibinfo{author}{\bibfnamefont{M.}~\bibnamefont{Hashimoto}},
  \bibinfo{author}{\bibfnamefont{D.}~\bibnamefont{Lu}}, \bibnamefont{et~al.},
  \bibinfo{journal}{arXiv preprint arXiv:2210.06653}
  (\bibinfo{year}{2022}{\natexlab{b}}).

\bibitem[{\citenamefont{Miao et~al.}(2022)\citenamefont{Miao, Zhang, Li,
  Fabbris, Said, Tartaglia, Yilmaz, Vescovo, Yin, Murakami
  et~al.}}]{lee-fege-arxiv22}
\bibinfo{author}{\bibfnamefont{H.}~\bibnamefont{Miao}},
  \bibinfo{author}{\bibfnamefont{T.~T.} \bibnamefont{Zhang}},
  \bibinfo{author}{\bibfnamefont{H.~X.} \bibnamefont{Li}},
  \bibinfo{author}{\bibfnamefont{G.}~\bibnamefont{Fabbris}},
  \bibinfo{author}{\bibfnamefont{A.~H.} \bibnamefont{Said}},
  \bibinfo{author}{\bibfnamefont{R.}~\bibnamefont{Tartaglia}},
  \bibinfo{author}{\bibfnamefont{T.}~\bibnamefont{Yilmaz}},
  \bibinfo{author}{\bibfnamefont{E.}~\bibnamefont{Vescovo}},
  \bibinfo{author}{\bibfnamefont{J.~X.} \bibnamefont{Yin}},
  \bibinfo{author}{\bibfnamefont{S.}~\bibnamefont{Murakami}},
  \bibnamefont{et~al.}, \bibinfo{journal}{arXiv preprint arXiv:2210.06359}
  (\bibinfo{year}{2022}).

\bibitem[{\citenamefont{Shao et~al.}(2022)\citenamefont{Shao, Yin, Belopolski,
  You, Hou, Chen, Jiang, Hossain, Yahyavi, Hsu et~al.}}]{chang-fege-arxiv22}
\bibinfo{author}{\bibfnamefont{S.}~\bibnamefont{Shao}},
  \bibinfo{author}{\bibfnamefont{J.-X.} \bibnamefont{Yin}},
  \bibinfo{author}{\bibfnamefont{I.}~\bibnamefont{Belopolski}},
  \bibinfo{author}{\bibfnamefont{J.-Y.} \bibnamefont{You}},
  \bibinfo{author}{\bibfnamefont{T.}~\bibnamefont{Hou}},
  \bibinfo{author}{\bibfnamefont{H.}~\bibnamefont{Chen}},
  \bibinfo{author}{\bibfnamefont{Y.-X.} \bibnamefont{Jiang}},
  \bibinfo{author}{\bibfnamefont{M.~S.} \bibnamefont{Hossain}},
  \bibinfo{author}{\bibfnamefont{M.}~\bibnamefont{Yahyavi}},
  \bibinfo{author}{\bibfnamefont{C.-H.} \bibnamefont{Hsu}},
  \bibnamefont{et~al.}, \bibinfo{journal}{arXiv preprint arXiv:2206.12033}
  (\bibinfo{year}{2022}).

\bibitem[{\citenamefont{Zhou et~al.}(2022)\citenamefont{Zhou, Yan, Fan, Wang,
  and Wan}}]{wan-fege-dmi-arxiv22}
\bibinfo{author}{\bibfnamefont{H.}~\bibnamefont{Zhou}},
  \bibinfo{author}{\bibfnamefont{S.}~\bibnamefont{Yan}},
  \bibinfo{author}{\bibfnamefont{D.}~\bibnamefont{Fan}},
  \bibinfo{author}{\bibfnamefont{D.}~\bibnamefont{Wang}}, \bibnamefont{and}
  \bibinfo{author}{\bibfnamefont{X.}~\bibnamefont{Wan}},
  \bibinfo{journal}{arXiv preprint arXiv:2211.15545}  (\bibinfo{year}{2022}).

\bibitem[{\citenamefont{Wu et~al.}(2023)\citenamefont{Wu, Hu, Wang, and
  Wan}}]{wan-fege-cdw-arxiv22}
\bibinfo{author}{\bibfnamefont{L.}~\bibnamefont{Wu}},
  \bibinfo{author}{\bibfnamefont{Y.}~\bibnamefont{Hu}},
  \bibinfo{author}{\bibfnamefont{D.}~\bibnamefont{Wang}}, \bibnamefont{and}
  \bibinfo{author}{\bibfnamefont{X.}~\bibnamefont{Wan}},
  \bibinfo{journal}{arXiv preprint arXiv:2302.03622}  (\bibinfo{year}{2023}).

\bibitem[{\citenamefont{Bernhardt and LebechS}(1984)}]{Bernhardt1984}
\bibinfo{author}{\bibfnamefont{J.}~\bibnamefont{Bernhardt}} \bibnamefont{and}
  \bibinfo{author}{\bibfnamefont{B.}~\bibnamefont{LebechS}},
  \bibinfo{journal}{J. Phys. F: Met. Phys} \textbf{\bibinfo{volume}{14}},
  \bibinfo{pages}{2379} (\bibinfo{year}{1984}).

\bibitem[{\citenamefont{Liechtenstein et~al.}(1995)\citenamefont{Liechtenstein,
  Anisimov, and Zaanen}}]{Liechtenstein1995}
\bibinfo{author}{\bibfnamefont{A.~I.} \bibnamefont{Liechtenstein}},
  \bibinfo{author}{\bibfnamefont{V.~I.} \bibnamefont{Anisimov}},
  \bibnamefont{and} \bibinfo{author}{\bibfnamefont{J.}~\bibnamefont{Zaanen}},
  \bibinfo{journal}{Physical Review B} \textbf{\bibinfo{volume}{52}},
  \bibinfo{pages}{R5467} (\bibinfo{year}{1995}), ISSN
  \bibinfo{issn}{0163-1829}.

\bibitem[{\citenamefont{Cococcioni and de~Gironcoli}(2005)}]{Cococcioni2005}
\bibinfo{author}{\bibfnamefont{M.}~\bibnamefont{Cococcioni}} \bibnamefont{and}
  \bibinfo{author}{\bibfnamefont{S.}~\bibnamefont{de~Gironcoli}},
  \bibinfo{journal}{Physical Review B} \textbf{\bibinfo{volume}{71}},
  \bibinfo{pages}{035105} (\bibinfo{year}{2005}), ISSN
  \bibinfo{issn}{1098-0121}.

\bibitem[{sup()}]{supp_info}
\bibinfo{note}{See Supplementary Information for 1. Tight-binding theory and
  tight-binding models of the 2$\times$2 structure. 2. Calculational method s.
  3. Discussions of the topological properties. 4. Bare and RPA generalized
  susceptibility tensors of the pristine FeGe. 5. Additional figures.}

\bibitem[{\citenamefont{Giustino}(2017)}]{epi-rmp17}
\bibinfo{author}{\bibfnamefont{F.}~\bibnamefont{Giustino}},
  \bibinfo{journal}{Rev. Mod. Phys.} \textbf{\bibinfo{volume}{89}},
  \bibinfo{pages}{015003} (\bibinfo{year}{2017}).

\bibitem[{\citenamefont{Liu and Balents}(2017)}]{liu-prb17}
\bibinfo{author}{\bibfnamefont{J.}~\bibnamefont{Liu}} \bibnamefont{and}
  \bibinfo{author}{\bibfnamefont{L.}~\bibnamefont{Balents}},
  \bibinfo{journal}{Phys. Rev. B} \textbf{\bibinfo{volume}{95}},
  \bibinfo{pages}{075426} (\bibinfo{year}{2017}).

\bibitem[{\citenamefont{Uehara et~al.}(2015)\citenamefont{Uehara, Shinaoka, and
  Motome}}]{motome-prb15}
\bibinfo{author}{\bibfnamefont{A.}~\bibnamefont{Uehara}},
  \bibinfo{author}{\bibfnamefont{H.}~\bibnamefont{Shinaoka}}, \bibnamefont{and}
  \bibinfo{author}{\bibfnamefont{Y.}~\bibnamefont{Motome}},
  \bibinfo{journal}{Phys. Rev. B} \textbf{\bibinfo{volume}{92}},
  \bibinfo{pages}{195150} (\bibinfo{year}{2015}).

\bibitem[{\citenamefont{Qiu et~al.}(2021)\citenamefont{Qiu, Zou, Luo, Cui,
  Song, Gao, Wang, and Xu}}]{xu-prl21}
\bibinfo{author}{\bibfnamefont{W.-X.} \bibnamefont{Qiu}},
  \bibinfo{author}{\bibfnamefont{J.-Y.} \bibnamefont{Zou}},
  \bibinfo{author}{\bibfnamefont{A.-Y.} \bibnamefont{Luo}},
  \bibinfo{author}{\bibfnamefont{Z.-H.} \bibnamefont{Cui}},
  \bibinfo{author}{\bibfnamefont{Z.-D.} \bibnamefont{Song}},
  \bibinfo{author}{\bibfnamefont{J.-H.} \bibnamefont{Gao}},
  \bibinfo{author}{\bibfnamefont{Y.-L.} \bibnamefont{Wang}}, \bibnamefont{and}
  \bibinfo{author}{\bibfnamefont{G.}~\bibnamefont{Xu}}, \bibinfo{journal}{Phys.
  Rev. Lett.} \textbf{\bibinfo{volume}{127}}, \bibinfo{pages}{147202}
  (\bibinfo{year}{2021}).

\bibitem[{\citenamefont{Ma et~al.}(2022)\citenamefont{Ma, Yin, Zahid~Hasan, and
  Liu}}]{ma-prb22}
\bibinfo{author}{\bibfnamefont{H.-Y.} \bibnamefont{Ma}},
  \bibinfo{author}{\bibfnamefont{J.-X.} \bibnamefont{Yin}},
  \bibinfo{author}{\bibfnamefont{M.}~\bibnamefont{Zahid~Hasan}},
  \bibnamefont{and} \bibinfo{author}{\bibfnamefont{J.}~\bibnamefont{Liu}},
  \bibinfo{journal}{Phys. Rev. B} \textbf{\bibinfo{volume}{106}},
  \bibinfo{pages}{155125} (\bibinfo{year}{2022}).

\bibitem[{\citenamefont{Teng et~al.}(2022{\natexlab{c}})\citenamefont{Teng,
  Chen, Ye, Rosenberg, Liu, Yin, Jiang, Oh, Hasan, Neubauer
  et~al.}}]{dai-fege-arxiv22}
\bibinfo{author}{\bibfnamefont{X.}~\bibnamefont{Teng}},
  \bibinfo{author}{\bibfnamefont{L.}~\bibnamefont{Chen}},
  \bibinfo{author}{\bibfnamefont{F.}~\bibnamefont{Ye}},
  \bibinfo{author}{\bibfnamefont{E.}~\bibnamefont{Rosenberg}},
  \bibinfo{author}{\bibfnamefont{Z.}~\bibnamefont{Liu}},
  \bibinfo{author}{\bibfnamefont{J.-X.} \bibnamefont{Yin}},
  \bibinfo{author}{\bibfnamefont{Y.-X.} \bibnamefont{Jiang}},
  \bibinfo{author}{\bibfnamefont{J.~S.} \bibnamefont{Oh}},
  \bibinfo{author}{\bibfnamefont{M.~Z.} \bibnamefont{Hasan}},
  \bibinfo{author}{\bibfnamefont{K.~J.} \bibnamefont{Neubauer}},
  \bibnamefont{et~al.}, \bibinfo{journal}{arXiv preprint arXiv:2203.11467}
  (\bibinfo{year}{2022}{\natexlab{c}}).

\bibitem[{\citenamefont{Kane and Mele}(2005)}]{Kane2005}
\bibinfo{author}{\bibfnamefont{C.~L.} \bibnamefont{Kane}} \bibnamefont{and}
  \bibinfo{author}{\bibfnamefont{E.~J.} \bibnamefont{Mele}},
  \bibinfo{journal}{Physical Review Letters} \textbf{\bibinfo{volume}{95}}
  (\bibinfo{year}{2005}), ISSN \bibinfo{issn}{00319007}.

\bibitem[{\citenamefont{Polshyn et~al.}(2022)\citenamefont{Polshyn, Zhang,
  Kumar, Soejima, Ledwith, Watanabe, Taniguchi, Vishwanath, Zaletel, and
  Young}}]{young-np22}
\bibinfo{author}{\bibfnamefont{H.}~\bibnamefont{Polshyn}},
  \bibinfo{author}{\bibfnamefont{Y.}~\bibnamefont{Zhang}},
  \bibinfo{author}{\bibfnamefont{M.~A.} \bibnamefont{Kumar}},
  \bibinfo{author}{\bibfnamefont{T.}~\bibnamefont{Soejima}},
  \bibinfo{author}{\bibfnamefont{P.}~\bibnamefont{Ledwith}},
  \bibinfo{author}{\bibfnamefont{K.}~\bibnamefont{Watanabe}},
  \bibinfo{author}{\bibfnamefont{T.}~\bibnamefont{Taniguchi}},
  \bibinfo{author}{\bibfnamefont{A.}~\bibnamefont{Vishwanath}},
  \bibinfo{author}{\bibfnamefont{M.~P.} \bibnamefont{Zaletel}},
  \bibnamefont{and} \bibinfo{author}{\bibfnamefont{A.~F.} \bibnamefont{Young}},
  \bibinfo{journal}{Nature Physics} \textbf{\bibinfo{volume}{18}},
  \bibinfo{pages}{42} (\bibinfo{year}{2022}), ISSN \bibinfo{issn}{1745-2481}.

\bibitem[{\citenamefont{Pierce et~al.}(2021)\citenamefont{Pierce, Xie, Park,
  Khalaf, Lee, Cao, Parker, Forrester, Chen, Watanabe et~al.}}]{pablo-np21}
\bibinfo{author}{\bibfnamefont{A.~T.} \bibnamefont{Pierce}},
  \bibinfo{author}{\bibfnamefont{Y.}~\bibnamefont{Xie}},
  \bibinfo{author}{\bibfnamefont{J.~M.} \bibnamefont{Park}},
  \bibinfo{author}{\bibfnamefont{E.}~\bibnamefont{Khalaf}},
  \bibinfo{author}{\bibfnamefont{S.~H.} \bibnamefont{Lee}},
  \bibinfo{author}{\bibfnamefont{Y.}~\bibnamefont{Cao}},
  \bibinfo{author}{\bibfnamefont{D.~E.} \bibnamefont{Parker}},
  \bibinfo{author}{\bibfnamefont{P.~R.} \bibnamefont{Forrester}},
  \bibinfo{author}{\bibfnamefont{S.}~\bibnamefont{Chen}},
  \bibinfo{author}{\bibfnamefont{K.}~\bibnamefont{Watanabe}},
  \bibnamefont{et~al.}, \bibinfo{journal}{Nature Physics}
  \textbf{\bibinfo{volume}{17}}, \bibinfo{pages}{1210} (\bibinfo{year}{2021}),
  ISSN \bibinfo{issn}{1745-2481}.

\bibitem[{\citenamefont{Qi et~al.}(2010)\citenamefont{Qi, Hughes, and
  Zhang}}]{qi-chiral-prb10}
\bibinfo{author}{\bibfnamefont{X.-L.} \bibnamefont{Qi}},
  \bibinfo{author}{\bibfnamefont{T.~L.} \bibnamefont{Hughes}},
  \bibnamefont{and} \bibinfo{author}{\bibfnamefont{S.-C.} \bibnamefont{Zhang}},
  \bibinfo{journal}{Phys. Rev. B} \textbf{\bibinfo{volume}{82}},
  \bibinfo{pages}{184516} (\bibinfo{year}{2010}).

\bibitem[{\citenamefont{Kresse and
  Furthm{\"u}ller}(1996)}]{kresse1996efficient}
\bibinfo{author}{\bibfnamefont{G.}~\bibnamefont{Kresse}} \bibnamefont{and}
  \bibinfo{author}{\bibfnamefont{J.}~\bibnamefont{Furthm{\"u}ller}},
  \bibinfo{journal}{Physcial Review B} \textbf{\bibinfo{volume}{54}},
  \bibinfo{pages}{11169} (\bibinfo{year}{1996}).

\bibitem[{\citenamefont{Perdew et~al.}(1996)\citenamefont{Perdew, Burke, and
  Ernzerhof}}]{perdew1996generalized}
\bibinfo{author}{\bibfnamefont{J.~P.} \bibnamefont{Perdew}},
  \bibinfo{author}{\bibfnamefont{K.}~\bibnamefont{Burke}}, \bibnamefont{and}
  \bibinfo{author}{\bibfnamefont{M.}~\bibnamefont{Ernzerhof}},
  \bibinfo{journal}{Physcial Review Letters} \textbf{\bibinfo{volume}{77}},
  \bibinfo{pages}{3865} (\bibinfo{year}{1996}).

\bibitem[{\citenamefont{Mostofi et~al.}(2008)\citenamefont{Mostofi, Yates, Lee,
  Souza, Vanderbilt, and Marzari}}]{mostofi2008wannier90}
\bibinfo{author}{\bibfnamefont{A.~A.} \bibnamefont{Mostofi}},
  \bibinfo{author}{\bibfnamefont{J.~R.} \bibnamefont{Yates}},
  \bibinfo{author}{\bibfnamefont{Y.-S.} \bibnamefont{Lee}},
  \bibinfo{author}{\bibfnamefont{I.}~\bibnamefont{Souza}},
  \bibinfo{author}{\bibfnamefont{D.}~\bibnamefont{Vanderbilt}},
  \bibnamefont{and} \bibinfo{author}{\bibfnamefont{N.}~\bibnamefont{Marzari}},
  \bibinfo{journal}{Computer physics communications}
  \textbf{\bibinfo{volume}{178}}, \bibinfo{pages}{685} (\bibinfo{year}{2008}).

\bibitem[{\citenamefont{Rath and Freeman}(1975)}]{interpolation-prb75}
\bibinfo{author}{\bibfnamefont{J.}~\bibnamefont{Rath}} \bibnamefont{and}
  \bibinfo{author}{\bibfnamefont{A.~J.} \bibnamefont{Freeman}},
  \bibinfo{journal}{Phys. Rev. B} \textbf{\bibinfo{volume}{11}},
  \bibinfo{pages}{2109} (\bibinfo{year}{1975}).

\bibitem[{\citenamefont{Georges et~al.}(2013)\citenamefont{Georges, Medici, and
  Mravlje}}]{georges2013strong}
\bibinfo{author}{\bibfnamefont{A.}~\bibnamefont{Georges}},
  \bibinfo{author}{\bibfnamefont{L.~d.} \bibnamefont{Medici}},
  \bibnamefont{and} \bibinfo{author}{\bibfnamefont{J.}~\bibnamefont{Mravlje}},
  \bibinfo{journal}{Annu. Rev. Condens. Matter Phys.}
  \textbf{\bibinfo{volume}{4}}, \bibinfo{pages}{137} (\bibinfo{year}{2013}).

\bibitem[{\citenamefont{Yin et~al.}(2019)\citenamefont{Yin, Zhang, Chang, Wang,
  Tsirkin, Guguchia, Lian, Zhou, Jiang, Belopolski et~al.}}]{Yin2019}
\bibinfo{author}{\bibfnamefont{J.-X.} \bibnamefont{Yin}},
  \bibinfo{author}{\bibfnamefont{S.~S.} \bibnamefont{Zhang}},
  \bibinfo{author}{\bibfnamefont{G.}~\bibnamefont{Chang}},
  \bibinfo{author}{\bibfnamefont{Q.}~\bibnamefont{Wang}},
  \bibinfo{author}{\bibfnamefont{S.~S.} \bibnamefont{Tsirkin}},
  \bibinfo{author}{\bibfnamefont{Z.}~\bibnamefont{Guguchia}},
  \bibinfo{author}{\bibfnamefont{B.}~\bibnamefont{Lian}},
  \bibinfo{author}{\bibfnamefont{H.}~\bibnamefont{Zhou}},
  \bibinfo{author}{\bibfnamefont{K.}~\bibnamefont{Jiang}},
  \bibinfo{author}{\bibfnamefont{I.}~\bibnamefont{Belopolski}},
  \bibnamefont{et~al.}, \bibinfo{journal}{Nature Physics}
  \textbf{\bibinfo{volume}{15}}, \bibinfo{pages}{443} (\bibinfo{year}{2019}),
  ISSN \bibinfo{issn}{1745-2473}.

\bibitem[{\citenamefont{Liu et~al.}(2018{\natexlab{b}})\citenamefont{Liu, Sun,
  Kumar, Muechler, Sun, Jiao, Yang, Liu, Liang, Xu et~al.}}]{Liu2018}
\bibinfo{author}{\bibfnamefont{E.}~\bibnamefont{Liu}},
  \bibinfo{author}{\bibfnamefont{Y.}~\bibnamefont{Sun}},
  \bibinfo{author}{\bibfnamefont{N.}~\bibnamefont{Kumar}},
  \bibinfo{author}{\bibfnamefont{L.}~\bibnamefont{Muechler}},
  \bibinfo{author}{\bibfnamefont{A.}~\bibnamefont{Sun}},
  \bibinfo{author}{\bibfnamefont{L.}~\bibnamefont{Jiao}},
  \bibinfo{author}{\bibfnamefont{S.-Y.} \bibnamefont{Yang}},
  \bibinfo{author}{\bibfnamefont{D.}~\bibnamefont{Liu}},
  \bibinfo{author}{\bibfnamefont{A.}~\bibnamefont{Liang}},
  \bibinfo{author}{\bibfnamefont{Q.}~\bibnamefont{Xu}}, \bibnamefont{et~al.},
  \bibinfo{journal}{Nature Physics} \textbf{\bibinfo{volume}{14}},
  \bibinfo{pages}{1125} (\bibinfo{year}{2018}{\natexlab{b}}), ISSN
  \bibinfo{issn}{1745-2473}.

\bibitem[{\citenamefont{Wang et~al.}(2018)\citenamefont{Wang, Xu, Lou, Liu, Li,
  Huang, Shen, Weng, Wang, and Lei}}]{Wang2018}
\bibinfo{author}{\bibfnamefont{Q.}~\bibnamefont{Wang}},
  \bibinfo{author}{\bibfnamefont{Y.}~\bibnamefont{Xu}},
  \bibinfo{author}{\bibfnamefont{R.}~\bibnamefont{Lou}},
  \bibinfo{author}{\bibfnamefont{Z.}~\bibnamefont{Liu}},
  \bibinfo{author}{\bibfnamefont{M.}~\bibnamefont{Li}},
  \bibinfo{author}{\bibfnamefont{Y.}~\bibnamefont{Huang}},
  \bibinfo{author}{\bibfnamefont{D.}~\bibnamefont{Shen}},
  \bibinfo{author}{\bibfnamefont{H.}~\bibnamefont{Weng}},
  \bibinfo{author}{\bibfnamefont{S.}~\bibnamefont{Wang}}, \bibnamefont{and}
  \bibinfo{author}{\bibfnamefont{H.}~\bibnamefont{Lei}},
  \bibinfo{journal}{Nature Communications} \textbf{\bibinfo{volume}{9}},
  \bibinfo{pages}{3681} (\bibinfo{year}{2018}), ISSN \bibinfo{issn}{2041-1723}.

\bibitem[{\citenamefont{Trumpy et~al.}(1970)\citenamefont{Trumpy, Both,
  Dj\'ega-Mariadassou, and Lecocq}}]{Trumpy-PRB-1970}
\bibinfo{author}{\bibfnamefont{G.}~\bibnamefont{Trumpy}},
  \bibinfo{author}{\bibfnamefont{E.}~\bibnamefont{Both}},
  \bibinfo{author}{\bibfnamefont{C.}~\bibnamefont{Dj\'ega-Mariadassou}},
  \bibnamefont{and} \bibinfo{author}{\bibfnamefont{P.}~\bibnamefont{Lecocq}},
  \bibinfo{journal}{Phys. Rev. B} \textbf{\bibinfo{volume}{2}},
  \bibinfo{pages}{3477} (\bibinfo{year}{1970}),
  \urlprefix\url{https://link.aps.org/doi/10.1103/PhysRevB.2.3477}.

\end{thebibliography}

\end{document}